\documentclass[reqno,12pt]{amsart} 
\usepackage{verbatim}

\usepackage{amsmath} 
\usepackage{amssymb} 
\usepackage[mathscr]{eucal}
\usepackage{graphicx}
\usepackage{graphics}
\usepackage{color}
\usepackage[normalem]{ulem}
\usepackage{bbold}

\numberwithin{equation}{section}

\usepackage{srcltx} 
\newcommand{\ignore}[1]{{}}

\def\Ascr{\mathscr{A}}
\def\Bscr{\mathscr{B}}
\def\Mscr{\mathscr{M}}

\newcommand{\abs}[1]{{\lvert #1\rvert}}

\newcommand{\norm}[1]{{\lVert #1\rVert}}

\def\cbX{\boldsymbol{\mathcal X}}

\def\cbV{\boldsymbol{\mathcal V}}

\def\cbM{\boldsymbol{\mathcal M}}

\def\bcdot{\boldsymbol{\cdot}}

\def\emptyset{\varnothing} 
\def\d{\delta} 
\def\e{{\rm e}} 
 
\def\L{\Lambda}

\newfam\Bbbfam 
\font\tenBbb=msbm10 
\font\sevenBbb=msbm7 
\font\fiveBbb=msbm5 
\textfont\Bbbfam=\tenBbb 
\scriptfont\Bbbfam=\sevenBbb 
\scriptscriptfont\Bbbfam=\fiveBbb

 \newcommand{\C}     {\mathbb{C}} 
\newcommand{\R}     {\mathbb{R}} 
\newcommand{\Z}     {\mathbb{Z}} 
\newcommand{\N}     {\mathbb{N}}

\newcommand{\E}     {\mathbb{E}} 
 
\newcommand{\T}     {\mathbb{T}}

\def\1{{\mathchoice {1\mskip-4mu\mathrm l}      % Blackboard bold 1 
{1\mskip-4mu\mathrm l} 
{1\mskip-4.5mu\mathrm l} {1\mskip-5mu\mathrm l}}} 
 
\def\comment#1{} 
\newtheoremstyle{thm}{2ex}{2ex}{\itshape\rmfamily}{} 
{\bfseries\rmfamily}{}{1.7ex}{} 
 
\newtheoremstyle{rem}{1.3ex}{1.3ex}{\rmfamily}{} 
{\itshape\rmfamily}{}{1.5ex}{} 
 
\newenvironment{proofsect}[1] 
{\vskip0.1cm\noindent{\bf #1.}\hskip0.5cm}

\newtheorem{theorem}{Theorem}[section] 
\newtheorem{lemma}[theorem]{Lemma} 
\newtheorem{prop}[theorem] {Proposition} 
 
\newtheorem{remark}[theorem]  {Remark}

\theoremstyle{definition}

\renewcommand{\d}{{\rm d}}

\newcommand{\id}{{\rm{id}}} 
 
\newcommand{\supp}{{\operatorname {supp}}}

\newcommand{\dist}{{\operatorname {dist}}}

\newcommand{\Exp}{\mathscr{E}\kern-0.2mm{\operatorname{xp}}}
\newcommand{\Log}{\mathscr{L}\kern-0.2mm{\operatorname{og}}}

\newcommand{\Acal}   {{\mathcal A }}
\newcommand{\Bcal}   {{\mathcal B }}
\newcommand{\Ccal}   {{\mathcal C }} 
 
\newcommand{\Ecal}   {{\mathcal E }}

\newcommand{\Kcal}   {{\mathcal K }} 
\newcommand{\Lcal}   {{\mathcal L }} 
\newcommand{\Mcal}   {{\mathcal M }}

\newcommand{\Rcal}   {{\mathcal R }} 
 
\newcommand{\Tcal}   {{\mathcal T }}

\newcommand{\m} {{\mathfrak m}}

\def\cbH{\boldsymbol{\mathcal H}}
\def\bcdot{\boldsymbol{\cdot}}
\def\Ascr{\mathscr{A}}
\def\Bscr{\mathscr{B}}
\def\Kscr{\mathscr{K}}
\def\Tscr{\mathscr{T}}
\def\Rscr{\mathscr{R}}
\def\Cscr{\mathscr{C}}
\def\Dscr{\mathscr{D}}

\def\aalpha{\alpha}

 \definecolor{green}   {cmyk}{0.91, 0   , 0.88, 0.12}

\begin{document} 
 
\title[Finite range decomposition]{\large 
Finite range decomposition for families of gradient Gaussian measures} 
 
\author[Stefan Adams, Roman Koteck\'{y}, Stefan M\"uller]{} 
\maketitle

\thispagestyle{empty} 
\vspace{0.2cm} 
 
\centerline {\sc By Stefan Adams\footnote{Mathematics Institute, University of Warwick, Coventry CV4 7AL, United Kingdom,\\ {\tt S.Adams@warwick.ac.uk}},  
Roman Koteck\'{y}\footnote{Center for Theoretical Study, Charles University, Prague, and Mathematics Institute, University of Warwick, United Kingdom, {\tt R.Kotecky@warwick.ac.uk}}\/ and Stefan M\"uller\footnote{Hausdorff Center for Mathematics
\& Institute for Applied Mathematics, Universit\"at Bonn,
Endenicher Allee 60, 53115 Bonn, Germany, {\tt stefan.mueller@hcm.uni-bonn.de}} }

\vspace{0.4cm}
\renewcommand{\thefootnote}{}

\bigskip

\begin{quote} 
{\small {\bf Abstract:} Let a family of gradient Gaussian vector fields  on $ \Z^d $ be given. We show the existence of a uniform finite range decomposition of the corresponding covariance operators, that is, the covariance operator can be written as a sum of covariance operators 
%%(Gaussian fields) 
whose kernels are supported within cubes  of   diameters $ \sim L^k $. In addition we prove natural  regularity for the subcovariance operators and we obtain regularity bounds as we vary within the given family of gradient Gaussian measures.}
\end{quote}

\vfill

\bigskip\noindent
{\it MSC 2000.} 

\medskip\noindent
{\it Keywords and phrases. Gradient Gaussian field, covariance, renormalisation, Fourier multiplier} 
 
\eject 
 
\setcounter{section}{0}

\section{Introduction}\label{S:FRDapp}

In this paper we construct a  finite range decomposition for a family of translation invariant gradient Gaussian fields on $ \Z^d $ ($d \geq 2$)
which depends real-analytically on the quadratic from that defines the Gaussian field. More precisely, we 
consider a large torus $(\Z/L^N \Z)^d$ and obtain a finite range decomposition with estimates 
that do not depend on  $N$. 
Equivalently, we show that the discrete Greens function $\Ccal_{A}$ of the (elliptic) translation invariant difference operator
$\Ascr = \nabla^* A \nabla$ can be written as a sum $\Ccal_{A} = \sum_k \Ccal_{A,k}$ of positive kernels $\Ccal_{A,k}$ which 
are supported in cubes of size $\sim L^k$ with natural estimates for their discrete derivatives $\nabla^{{\aalpha}} \Ccal_{A,k}$ (see Theorem~\ref{THM:FRD})  as well as for their derivatives with respect to $A$ (see Theorem~\ref{THM:FRDfamily}).

To put this into perspective recall that an $\R^m$-valued  Gaussian field $\xi$ on $\Z^d$   (with vanishing expectation, $\E(\xi(x)) = 0$) is said to have range $M$ if
the correlation matrices $\E\big[\xi^r(x)\xi^s(y)\big]$, $r,s=1,\ldots, m, $ vanish whenever $|x - y| > M$. In the following  we consider
only translation invariant Gaussian fields. We say that Gaussian fields
$\xi_k$ form a finite range decomposition of $\xi$ if $\xi = \sum_k \xi_k$ and $\xi_k$ has range $\sim L^k$, where $L \geq 2$ is an integer. 
The existence of such a decomposition is equivalent to a decomposition of the correlation matrices 
$\Ccal_A(x,y)^{r,s} := \E\big[\xi^r(x)\xi^s(y)\big]$ as a sum of positive (semi-) definite matrix valued kernels $\Ccal_{A,k}$
with range $ \sim L^k$, i.e.,  $\sum_{x,y}\sum_{r,s} \Ccal_{A,k}^{r,s}(x,y) \xi^r(y) \xi^s(x)  \geq 0$ and $\Ccal_{A,k}(x,y) = 0$ if $|x -y| 
\gtrsim L^k$. 

We are interested in gradient Gaussian fields, i.e., Gaussian fields 
with $\sigma$-algebra determined by the gradients $\nabla \xi$.
Such fields
arise naturally e.g. in problems in elasticity where only the difference 
of values is relevant for the energy. In this case we seek a decomposition into gradient Gaussian fields 
such that the gradient-gradient correlation $\E\big[\nabla_i\xi^r(x) \nabla_j \xi^s(y)\big] = \nabla_i \nabla_j^* E\big[\xi^r(x)\xi^s(y)\big]$  
vanishes for $|x-y| \gtrsim L^k$. Gradient Gaussian fields are more subtle to handle since they exhibit long-range
correlations  (the gradient-gradient correlation of the original field typically has only  algebraic decay $|x-y|^{-d}$ with the critical
exponent $-d$). In the language of quantum field theory gradient Gaussian  fields are thus often  referred to as massless fields. 
 
Decomposition into a 
sum of positive definite operators has been discussed in \cite{HS02} where a radial function is written as a weighted integral of tent functions. In \cite{BGM04}, finite range decompositions of the resolvent of the Laplacian $ (a-\Delta)^{-1} $, with $ a\ge 0 $, have been obtained  both
for the usual   Laplacian and for  finite difference Laplacian on the simple cubic lattice $ \mathbb{Z}^d $. In \cite{BT06} these results are extended and
  generalised  by providing sufficient conditions for a positive definite function to admit decomposition into a sum of positive 
  functions that are compactly supported within disks of increasing diameters $ \frac{1}{2} L^k $. More precisely, the authors  of \cite{BT06} consider 
  positive definite bilinear forms on $ \Ccal_0^\infty $ and prove that finite range decompositions do exist when the bilinear form is dual to a bilinear form 
  $ \varphi\mapsto \int|B\varphi(x) |^2\,\d x $ where $ B $ is a vector valued partial differential operator satisfying some regularity conditions.

  The main novelty of our paper is twofold. First we extend the finite range decomposition for the discrete Laplacian to 
  a situation where no maximum principle is available (even in the scalar case there is no discrete maximum principle for general  elliptic difference operator $\nabla^* A \nabla$ with constant coefficients). This can be seen as an adaptation of \cite{BT06}  to the discrete setting.  Secondly, we show that the finite range decomposition can be chosen so that the kernels $\Ccal_{A,k}$ depend
  analytically on $A$ as long as $A$ is positive definite.

  Our main motivation is the renormalization group (RG) approach to problems in statistical mechanics, following the  longstanding 
  research programme 
  of Brydges and Yau \cite{BY90},  and in particular the  recent work of Brydges  \cite{Bry09}, both inspired by the work of 
K.G.~Wilson \cite{Wilson}. The goal is to get good control of the expectations $\E(K)$ 
of  nonlinear functions that    
depends on a  gradient Gaussian field $ \xi$ in a large region $ \L\subset\Z^d $ of the integer lattice.  
The size of $ \L $ and the long range correlations in $\xi$ make it difficult to obtain accurate estimates on the expectation $ \E(K) $. 
In  \cite{AKM12} we show that such control can nonetheless be obtained in many interesting cases using the RG approach. 
One key difference with  the 
  earlier work of Brydges and others is   the necessity to drop the assumption of   isotropy. Hence the 
relevant quadratic term is a general finite difference operator $\nabla^* A \nabla$ and reduces no longer to 
a multiple of the discrete Laplacian. For this reason we need a finite-range decomposition for general (elliptic) operators $A$ and, in addition, 
we need to control derivatives  of the finite range decomposition with respect to  $A$.

In Section~\ref{sec-results} we introduce  the setting of gradient fields and the relevant Greens functions.
Our two main results are given in Theorem~\ref{THM:FRD} and Theorem~\ref{THM:FRDfamily}.
The existence of the finite range decomposition is proved in Section~\ref{sec-mainproof} where we adapt and 
extend the methods in \cite{BT06} to our setting. The regularity estimate for  a fixed $A$  is established in Section \ref{sec-mainproofestimates}.
Real-analytic dependence on $A$ is proved in Section \ref{sec-mainproofanalytic}.

\subsection*{Acknowledgments}
The research of S.A. was supported by EPSRC grant  EP/I003746/1,
R.K.  by the grants GA\v CR 201-09-1931, 201/12/2613, and MSM 0021620845, and S.M.  by the DFG Research group 718 ``Analysis and Stochastics in Complex Physical Systems''.

%%%%%

\section{Notation and main results}\label{sec-results}
We are interested in  gradient Gaussian fields on bounded domains in $ \Z^d $. For that let $ L \geq 3 $ be a fixed  odd 
integer and consider for any integer $N$ the space 
$$
 \cbV_N=\{\varphi: \mathbb Z^d\to\mathbb R^m;\  \varphi(x+ z)=\varphi(x)\  \text{ for all }\  z\in (L^N\mathbb Z)^d\}
=\bigl(\R^m\bigr)^{\T_N}
$$
of functions on the  torus
$\mathbb \T_N:=\bigl(\mathbb Z/L^N\mathbb Z\bigr)^d$
equipped with with the scalar product
\begin{equation}
\langle\varphi,\psi\rangle=\sum_{x\in \mathbb \T_N}\langle\varphi(x),\psi(x)\rangle_{\R^m}.
\end{equation}
Notice that a function on $\T_N$ can be identified with an $L^N$-periodic function on $\Z^d$.
We will later denote the corresponding space of $\C^m$-valued function, equipped with the usual scalar product
in the same way.
 
We consider two  distances  on $\Z^d$:  $\rho(x,y)=\inf\{\abs{x-y+z}\colon z\in (L^N\mathbb Z)^d\}$ and $\rho_\infty(x,y)=
\inf\{\abs{x-y+z}_\infty\colon  z\in (L^N\mathbb Z)^d\}$. Then the torus  can be represented  by the lattice cube 
$\T_N =\{x\in\Z^d\colon \abs{x}_\infty \le \frac{1}{2}(L^N-1)\} $ of side $ L^N $, equipped with the metric $\rho$ or $\rho_\infty$. 
Gradient Gaussian fields can be easily defined as discrete gradients  of Gaussian fields. However it turns
out to be inconvenient to work directly with the space of discrete gradient fields, since the constraint of being curl free (in a discrete sense)
leads to  a complicated bookkeeping. Instead, we use that discrete gradient fields are in one-to-one relation to usual fields modulo a constant.
%
%impose technical challenges which we circumvent defining a Gaussian fields with additional constraint, i.e., we shall delete a constant the obtain a bijection from the field to its gradient. 
To eliminate this constant we  use the normalisation condition that the sum of the field  over the torus vanishes. We thus denote by
 $ \cbX_N $ the subspace
\begin{equation}
\cbX_N=\{\varphi\in \cbV_N: \sum_{x\in \mathbb T_N }\varphi(x)=0\}.
\end{equation}

The forward and backward derivatives are defined as
\begin{equation}
\begin{aligned}
(\nabla \varphi)_j^{r}(x)&=\varphi^{ r}(x+{\rm e}_j)-\varphi^{r}(x),\\
(\nabla^*\varphi)_j^{r}(x)&=\varphi^{r}(x-{\rm e}_j)-\varphi^{r}(x), \quad {r}=1,\ldots, m; \ j=1,\ldots,d.
\end{aligned}
\end{equation}
Let $ A\colon\R^{m\times d}\to\R^{m\times d} $ be a linear map that is symmetric with respect to the standard scalar product $ (\cdot,\cdot)_{\R^{m\times d}} $ on $ \R^{m\times d} $ and positive definite, that is, there exists a constant $ c_0>0 $ such that
\begin{equation}
\label{E:AFF}
(AF,F)_{\R^{m\times d}}\ge c_0\norm{F}_{\R^{m\times d}}^2\quad\mbox{ for all } F\in\R^{m\times d}\mbox{ with } \norm{F}_{\R^{m\times d}}=(F,F)_{\R^{m\times d}}^{1/2}.
\end{equation}
The corresponding Dirichlet form defines a scalar product on $ \cbX_N $, 
\begin{equation}
\label{E:plusprod}
(\varphi,\psi)_+:=\Ecal(\varphi,\psi)=\sum_{x\in \T_N} \langle A(\nabla\varphi(x)),\nabla\psi(x)\rangle_{\R^{m\times d}}, \quad \varphi,\psi\in\cbX_N.
\end{equation}

Skipping the index $N$, we consider the triplet  $\cbH_- = \cbH   =  \cbH_+$  of (finite-dimensional) Hilbert spaces obtained by equipping the space  $\cbX_N$ with
 the norms $\norm{\cdot}_-$,  $\norm{\cdot}_2$,  and $\norm{\cdot}_+$, respectively.
 Here, $\norm{\cdot}_2 $ denotes the $ \ell_2 $-norm $\norm{\varphi}_2 = \langle\varphi,\varphi\rangle^{1/2}$,
 $\norm{\varphi}_+= (\varphi,\varphi)_+^{1/2}$, and $\norm{\cdot}_-$ is the dual norm \begin{equation}
\norm{\varphi}_-=\sup_{\psi: \norm{\psi}_+\le 1} \langle\psi, \varphi\rangle.
\end{equation}
One easily checks that $\norm{ \cdot }_-$ is again induced in a unique way by a scalar product $( \cdot, \cdot)_-$. The linear map $A$ defines an isometry
\begin{equation}
\label{Loperator}
\begin{aligned}
\Ascr\colon \cbH_+\to\cbH_-,\quad \varphi\mapsto \Ascr\varphi=\nabla^*(A\nabla\varphi).
\end{aligned}
\end{equation}
Indeed, it follows from the Lax-Milgram theorem that, for each $f \in \cbH_-$,  the equation 
\begin{equation}
(\varphi, v)_+ = \langle f, v \rangle \  \text{ for all }\   v \in \cbH_+
\end{equation}
has a unique solution  $\varphi \in \cbH_+$. Hence $\Ascr$ is a bijection from $\cbH_+$ to $\cbH_-$.
Moreover
\begin{equation}
\|\Ascr \varphi \|_-  = \sup \{ \langle \Ascr \varphi, v \rangle  : \|v\|_+ \leq 1 \}
= \sup \{(\varphi, v)_+ : \|v \|_+ \leq 1 \}  = \|\varphi\|_+  .
\end{equation}
Thus $\Ascr$ is an isometry from $\cbH_+$ to $\cbH_-$. In view of the symmetry of $\Ascr$ it  follows that
\begin{equation}
(\varphi, \psi)_- = (\Ascr^{-1} \varphi, \Ascr^{-1} \psi)_+
= \langle \Ascr^{-1} \varphi, \Ascr \Ascr^{-1} \psi \rangle 
= \langle \Ascr^{-1} \varphi,  \psi \rangle. 
\end{equation}

In the more abstract construction of 
Brydges and Talarczyk \cite{BT06} it is important that the operator $\Ascr$ can be written
as 
\begin{equation}\label{L1}
\Ascr={\Bscr}^*{\Bscr},
\end{equation} 
where $\Bscr^*$ denotes the dual of $\Bscr$.
This is indeed possible in our case. 
Since the operator $ A $  is symmetric and positive definite it  has a positive square root $ A^{1/2} $ and 
we can define
 $\Bscr$ by
\begin{equation}
\label{E:Bscr}
(\Bscr\varphi)(x)=(A^{1/2}\nabla\varphi)(x).
\end{equation}
This yields
\begin{equation}
\label{E:()_+}
(\varphi,\psi)_+=\langle\Bscr\varphi,\Bscr\psi\rangle,  \quad \norm{\varphi}_+=\norm{\Bscr\varphi}_2 .
\end{equation}
In the following, however, we will not use the operator $\Bscr$ explicitly and will, instead, directly use that 
 $\langle \Ascr \varphi, \psi \rangle  = \langle A \nabla \varphi, \nabla \psi \rangle$ and exploit
that the right hand side is sufficiently local in $\varphi$ and $\psi$. 
We do, however, make crucial use of the assumption that $A$ is positive definite in
the proof of Lemma \ref{lem2}.  For many of the other  estimates it would be sufficient to assume
that the operator $\Ascr$ is positive which is implied by the weaker condition that $A$ is positive
definite on matrices of rank one, i.e., 
$\langle A (a \otimes b), a \otimes b \rangle \ge c_0 |a|^2 |b|^2$ for all $a \in \R^m$, $b \in \R^d$.

Consider now the inverse $\Cscr_A={\Ascr}^{-1}$ of the operator $\Ascr$ (or
the Green function) and the corresponding bilinear form on $ \cbX_N $ defined by
\begin{equation}
\label{E:Gdef}
G_A(\varphi,\psi)=\langle\Cscr_A\varphi,\psi\rangle=(\varphi,\psi)_-,\quad \varphi,\psi\in\cbX_N.
\end{equation}
Given that the operator $\Ascr$  and its inverse commutes with translations on $\T_N$,
there exists a unique kernel  $\Ccal_A$ 
such that
\begin{equation}
(\Cscr_A\varphi)(x)=\sum_{y\in\T_N}\Ccal_A(x-y)\varphi(y),
\end{equation}
(see Lemma \ref{lemma:representation} below).  We write $\Ccal_A \in \cbM_N$, using  $\cbM_N$ 
(in analogy with $\cbX_N$) to denote  the space 
of all matrix-valued  maps on $\T_N$ with zero mean.
Notice that if the kernel $\Ccal_A$ is constant, $\Ccal_A(x)=C$ for any $x\in\T_N$, where $C$ is a linear operator on $\R^m$, then $(\Ccal_A\varphi)(x)= \sum_{y\in\T_N}\Ccal_A(x-y)\varphi(y)=C\sum_{y\in\T_N}\varphi(y)=0$
for any $\varphi\in\cbX_N$.
It is easy to see that  the function $G_{A,y}(\bcdot)=\Ccal_A(\bcdot-y)$ is the unique solution $G_{A,y}\in  \cbM_N$ of the equation
\begin{equation}
\Ascr G_{A,y}=\bigl(\delta_y -\frac1{L^{Nd}}\bigr) \mathbb 1,
\end{equation}
where $\mathbb 1$ is the unit $m\times m$ matrix.
Notice that for any $a\in\R^m$ one has:
$$
(\Ascr G_{A,y}) a=\bigl(\delta_y -\frac1{L^{Nd}}\bigr) a \in \cbX_N.
$$

We now state our main results.

\begin{theorem}\label{THM:FRD}
The operator $\Cscr_A \colon \cbH_-\to\cbH_+$  admits a finite range decomposition, i.e., there exist translation invariant positive-definite operators
 \begin{equation}
\Cscr_{A,k} \colon \cbH_-\to\cbH_+,\  (\Cscr_{A,k}\varphi)(x)=\sum_{y\in\T_N}\Ccal_{A,k}(x-y)\varphi(y),\     k=1,\dots, N+1,
\end{equation}
such that 
\begin{equation}\label{sumfrd}
\Cscr_A=\sum_{k=1}^{N+1} \Cscr_{A,k},
\end{equation}
and for  each associated kernel $\Ccal_{A,k} \in \cbM_N$ there exists a constant matrix $C_{A,k}$ such that
\begin{equation}\label{finiterange}
\Ccal_{A,k}(x-y)=  C_{A,k}\ \text{ whenever } \   \rho_\infty(x,y)\ge \frac{1}{2} L^k\quad \mbox{ for } k=1,\dots,N  .
\end{equation}
Moreover,  for any multiindex $ \aalpha $, there exist  constants  $ C_{\aalpha}(d)>0 $ and $\eta(\aalpha, d)$, depending only on the dimension $d$, such that
\begin{equation}\label{regularityboundfield}
\norm{\nabla^{\aalpha}\Ccal_{A,k}(x)}\le C_{\aalpha}(d) L^{-(k-1)(d-2+|\aalpha|)}
L^{\eta(\aalpha, d)}
\end{equation} for all $ x\in\T_N $ and $ k=1,\ldots,N+1 $. Here, $ \nabla^{\aalpha}=\prod_{i=1}^d\nabla_i^{\alpha_i} $
and $\nabla_i^0 = \id$, and $ \norm{\cdot} $ denotes the operator norm.
\end{theorem}

Note that since any function in $\cbX_N$ has mean zero the kernel $\widetilde{\Ccal_{A,k}} = \Ccal_{A,k} -C_{A,k}$ generates the same operator
$\Cscr_{A,k}$. Thus \eqref{finiterange} indeed guarantees that $\Cscr_{A,k}$ has finite range. See Lemmas~\ref{lemma:locality} 
and~\ref{lemma:representation} for further details.

The operator $\Ascr$, its inverse $\Cscr_{A}$, and the finite range decomposition  itself, depend on the linear map $A \colon \R^{m \times d} \to 
\R^{m \times d}$.  Our major result is that the finite range decomposition can be defined in such a way that the maps
$A \mapsto \Ccal_{A,k}$ are real-analytic, as long as $A$ is positive definite. 

Let $\Lcal_{\rm sym}(\R^{m\times d})$ denote the space of linear maps $A \colon \R^{m \times d} \to 
\R^{m \times d}$ that are symmetric with respect to the standard scalar product on $\R^{m \times d}$ and
let 
\begin{equation}
U := \left\{   A \in \Lcal_{\rm sym}(\R^{m\times d})\colon (A F, F)_{\R^{m\times d}} > 0 \ \text{ for all }\ F \in \R^{m \times d}, F\neq 0 \right\}
\end{equation}
denote the open subset of positive definite symmetric maps.

\begin{theorem}\label{THM:FRDfamily} Let $d \geq 2$ and let ${\aalpha}$ be a multiindex. There exist constants
$C_{\aalpha}(d)$ and $\eta(\aalpha, d)$ with the following properties. For each  integer $N \geq 1$, each $k=1, \dots, N + 1$ and each odd integer $L \geq 16$
there exist real-analytic maps  $A \mapsto \Ccal_{A,k}$ from $U$ to $\cbM_N$ such that the following three assertions hold.
\begin{enumerate}
\item[(i)] If $\Cscr_{A,k}$ denotes the translation invariant operator on induced by $\Ccal_{A,k}$ then
\begin{equation}
 \Cscr_{A} = \sum_{k=1}^{N+1} \Cscr_{A,k}.
 \end{equation}
\item[(ii)] There exist constant $m\times m$ matrices $C_{A,k}$ such that 
\begin{equation}
\Ccal_{A,k}(x) = C_{A,k}  \quad \mbox{if   } \rho_\infty(x,0) \ge \frac12  L^k.
\end{equation}
\item[(iii)] If $(A_0 F, F)_{\R^{m\times d}} \geq  c_0 \norm{F}_{\R^{m\times d}}^2$ for all $F \in \R^{m \times d}$ and $c_0 > 0$ then
\begin{equation}
\label{eq:analyticbounds}
\sup_{\norm{\dot{A}}\le 1}\Big\|\big(\nabla^{\aalpha}D_A^j\Ccal_{A_0,k}(x)(\dot{A},\ldots,\dot{A})\Big\|
\le C_{\aalpha}(d) \left(\frac{2}{c_0}\right)^j  j! \, L^{-(k-1)(d-2+|\aalpha|)}L^{\eta(\alpha, d)}.
\end{equation}
for all $ x\in\T_N $ and all $j \geq 0$.
%and all $ k=1,\ldots, N+1$ and all $ A\in M $
Here $ \nabla^{\aalpha}=\prod_{i=1}^d\nabla_i^{\alpha_i}$, we use
$\norm{\dot{A}} $ to denote the operator norm of a linear mapping $ \dot{A}\colon\R^{m\times d}\to\R^{m\times d}$, and the $j$th derivative with respect to $A$ in the direction $\dot{A}$ is taken at  $A_0$.
\end{enumerate}
\end{theorem}

The condition $L \geq 16$ can be dropped. However, in applications to renormalization group arguments, one needs to choose $L$ large
also for other reasons. Given $\delta \in (0, 1/2)$, the property (ii) can be strengthened to
$\Ccal_{A,k}(x) = C_{A,k}  \quad \mbox{if   } \rho_\infty(x,0) > \delta L^k$, provided that $L$ is large enough.  Then the constants
$C_{\aalpha}(d)$ and $\eta(\aalpha, d)$ depend also on $\delta$. We refer to Remark \ref{rem:smallrange} for further details. 

The proof of the existence of the finite range decomposition, i.e., \eqref{sumfrd} and \eqref{finiterange}  of Theorem~\ref{THM:FRD}, 
 is given in Section~\ref{sec-mainproof}. The remaining proof of the regularity bounds is given in Section~\ref{sec-mainproofestimates} for a fixed $A$. 
 Real-analytic dependence on $A$ and the bounds \eqref{eq:analyticbounds} are established in Section~\ref{sec-mainproofanalytic}.

The research of R.K. was partially supported by the grants GA\v CR 201-09-1931, 201/12/2613, and MSM 0021620845.

%\section{Proof of Theorem~\ref{THM:FRD}}•
\section{Construction of the finite range decomposition}
\label{sec-mainproof}

In this section we prove the existence part of Theorem~\ref{THM:FRD} via an extension and adaption of the methods in \cite{BT06} to our case. The proof of the estimates is  given in Sections~\ref{sec-mainproofestimates} and \ref{sec-mainproofanalytic}.

\noindent The existence of a finite range decomposition is contained
 in Proposition~\ref{P:FRDprop1} and Proposition~\ref{P:FRDprop2}  below and their  proofs are built  on the following auxiliary results.

First, for the construction of the decomposition we consider the discrete cube
\begin{equation}
Q=\{1,\ldots,l-1\}^d
\end{equation}
for some  $ l\in\N, l\ge 3 $. 
We can identify $Q$  with a subset of $\T_N$ once $l-1< L^N$. Similarly, any shift $Q+x\subset \T_N$.
%% For l =2  the cube Q  contains only one point and hence $\cbH_(Q+x)$ contains only the zero function
For any $ x\in  \T_N$, consider the subspace
\begin{equation}
 \cbH(Q+x)=\{\varphi\in \cbH\colon \varphi=0 \,\mbox{ in }\,\T_N\setminus (Q+x)\}.
\end{equation}
We write   $\cbH_+(Q+x)$ and $\cbH_-(Q+x)$ for the same space equipped with the scalar
products $( \cdot, \cdot)_+$ and $(\cdot, \cdot)_-$, respectively.
We denote by $ \varPi_x $ the $ (\cdot,\cdot)_+ $-orthogonal projection $\cbH_+\rightarrow \cbH_+(Q+x) $
and set  $P_x=\id- \varPi_x$. Thus $\varPi_x \varphi \in \cbH_+(Q+x)$ and
\begin{equation}  \label{eq:define_varpi}
(\varPi_x \varphi, \psi)_+  = (\varphi, \psi)_+  \  \text{ for all }\   \psi \in \cbH_+(Q+x)  .
\end{equation}
For any set $M \subset \Lambda_N$, we define its closure
by 
\begin{equation}
\overline M=\{x\in \Lambda_N\colon \dist_\infty(x,M)\le 1\}, \quad \dist_\infty(x,M) := \min \{ \rho_\infty(x,y)\colon y \in M \}.
\end{equation}
In particular,
\begin{equation}
\overline Q = \{0, \ldots, l\}^d.
\end{equation}
We also define
\begin{equation}
Q_- :=\{0,1,\ldots,l-1\}^d.
\end{equation}

%%
%notice that $P_x\varphi$ is a modified $\Ascr$-harmonic extension of the function $\varphi$ to $Q+x$ in the following sense.

\begin{lemma} \label{lemma:projection}
For any $\varphi\in \cbH_+$ we have
\begin{enumerate}
%\item[(i)] $P_x \varphi\in \cbH_+$,
%\smallskip
%
\item[(i)] \ \ $\Ascr(P_x \varphi) = \text{\rm const}$   in $Q+x$, 
\smallskip

\item[(ii)] \ \ $P_x \varphi=\varphi $ in $\T_N\setminus (Q+x)$,
\smallskip
%%\item[(iv)] If $\varphi = \psi$ in $(\bar Q \setminus Q) + x$ then $P\varphi = P\psi$ in $\bar{Q} + x$.
\item[(iii)]\ \ 
$\varPi_x \varphi = \varphi  1_{Q+x}  \quad  \mbox{if   }  \varphi = 0 \mbox{  on  }  \overline{(Q +x)} \setminus {(Q+x)}$.
\end{enumerate}
\end{lemma}

\begin{remark}  \label{rem:lowpasshighpass}
This shows that $P_x \varphi$ is essentially the $\Ascr$-harmonic extension in $Q+x$. Thus we would expect that
$P_x$ is (locally) smoothing and suppresses locally high frequency oscillations, while $\varPi_x$ suppresses locally low frequencies.
This will be made precise in Lemma \ref{Fouriermultipliers} below where we show the corresponding estimates for the averaged operators
$\Tscr = l^{-d} \sum_{ x \in \T_N} \varPi_x$ and $\Rscr = \id - \Tscr$. 
\end{remark}

\begin{proofsect}{Proof}
(i): By (\ref{eq:define_varpi}) we have for all  $\psi\in \cbH_+(Q+x)$  the relation $(P_x \varphi, \psi)_+ = 0$ and hence
$\langle\Ascr(P_x\varphi),\psi\rangle= 0$. Taking $\psi=\delta_v-\delta_z$ for any pair of points $v,z\in Q+x$ we get $\Ascr( P_x  \varphi)(v)=\Ascr(P_x  \varphi)(z)$. This proves (i). \\
(ii):  This follows from the fact that $\varPi_x \varphi$ belongs to $\cbH_+(Q+x)$ and hence vanishes outside $Q +x$. \\
(iii):  It suffices to consider the case $x= 0$ and we write $\varPi$ for $\varPi_0$. Let $\tilde \varphi =  \varphi 1_Q$. Then $\tilde \varphi \in \cbH_+(Q)$ and hence $\varPi \tilde \varphi = \tilde \varphi$. Moreover
$\varphi - \tilde \varphi$ vanishes in $\overline Q$. Thus $\nabla (\varphi - \tilde \varphi)$ vanishes in $Q_-$.
Hence $(\varphi - \tilde \varphi, \psi)_+ = 0$ for all $\psi \in \cbH_+(Q)$ since $\nabla \psi$ is supported in $Q_-$.
Therefore $\varPi (\varphi - \tilde \varphi) = 0$
which yields the assertion.  \\
\qed
\end{proofsect}

\begin{lemma}\hfill
\label{L:PixPiy}
\begin{enumerate}
\item[(i)] \ \ $
\varPi_x\varPi_y=0\,\mbox{ whenever }\, (Q_-+x)\cap(Q_-+y)=\emptyset $, 
\item[(ii)] \ \ $\varPi_x \varphi =0$ whenever $\supp \varphi \cap (\overline Q + x) = \emptyset$.
\end{enumerate}
\end{lemma}
\begin{proofsect}{Proof}
(i):  For any $ \varphi,\psi\in\cbH_+$,  the functions $ \varPi_x\varphi$ and  $ \varPi_y\psi$ vanish on $ \T_N\setminus (Q+x) $ and  $  \T_N\setminus(Q+y) $, respectively. Hence,  $\nabla \varPi_x \psi $ and $\nabla \varPi_y \varphi$ vanish on $ \T_N\setminus (Q_-+x) $ and  on $  \T_N\setminus(Q_-+y) $, respectively. Assuming now that 
 $Q_-+x $ and $Q_-+y $ are disjoint   and taking into account  (\ref{E:plusprod}) we get
\begin{equation}
\label{E:pixpiy}
(\psi,\varPi_x\varPi_y\varphi)_+=(\varPi_x\psi,\varPi_y\varphi)_+= 
\sum_{ z \in \T_N} \langle A( \nabla \varPi_x\psi)(z) , (\nabla \varPi_y\varphi)(z) \rangle_{\R^{m\times d}}=0 .
\end{equation}

\noindent (ii): For $\psi \in \cbH_+(Q+x)$ we have $\Ascr \psi = 0$ in $ \T_N \setminus (\overline Q + x)$. Thus for any $ \varphi \in  \cbH_+$ 
with $\supp \varphi \cap (\overline{Q} + x) = \emptyset$ we get
$(\varphi, \psi)_+ = \langle \varphi, \Ascr \psi \rangle = 0$.   In view of (\ref{eq:define_varpi}) this yields $\varPi_x \varphi = 0$. 
\qed
\end{proofsect}

\noindent Next, consider the symmetric operator
\begin{equation}
\label{E:T}
\Tscr=\frac{1}{l^d}\sum_{x\in \T_N}\varPi_x
\end{equation}
on $\cbH_+$.
The following result is the key estimate for the finite range decomposition construction. Our proof is a slight modification 
of the argument in \cite{BT06}.

\begin{lemma} \label{lem2} 
For any $ \varphi\in\cbH_+ $ we have
\begin{enumerate}
\item[(i)]\ \
$0 \leq (\varPi_x \varphi, \varphi)_+ \leq \langle  1_{Q_- + x} A \nabla \varphi, \nabla \varphi \rangle,$
\item[(ii)]\ \
$0 \leq (\Tscr \varphi, \varphi)_+ \leq (\varphi, \varphi)_+$
and the inequalities are strict if $\varphi \neq 0$, 
\item[(iii)]\ \
$(\Tscr \varphi, \Tscr \varphi)_+ \leq (\Tscr \varphi, \varphi)_+$ .
\end{enumerate}
\end{lemma}

\begin{proofsect}{Proof}
(i): We have  $(\varPi_x \varphi, \varphi)_+ =  ( \varphi, \varPi_x \varphi)_+ =  (\varPi_x \varphi, \varPi_x \varphi)_+ \geq 0$. 
For the other inequality we use that $\nabla \varPi_x \varphi$ is supported in $Q_- +x$. Thus
\begin{equation}   \label{eq:lemma2a}
 (\varPi_x \varphi, \varphi)_+  = \langle A \nabla \varPi_x \varphi, \nabla \varphi \rangle
 = \langle  A \nabla \varPi_x \varphi,  1_{Q_- +x }\nabla \varphi \rangle.
\end{equation}
Since $A$ is symmetric and positive definite the expression $(F,G)_A := \langle AF, G \rangle$ is a scalar product
on functions $\Z^d\to \R^{m \times d}$. Thus the Cauchy-Schwarz inequality yields
\begin{eqnarray}  \label{eq:lemma2b}
\langle  A \nabla \varPi_x \varphi,  1_{Q_- +x }\nabla \varphi \rangle 
&\leq&  \langle  A \nabla \varPi_x \varphi,  \nabla \varPi_x \varphi \rangle^{1/2}
\langle  A  1_{Q_- + x} \nabla  \varphi,  1_{Q_- +x }\nabla \varphi \rangle^{1/2}  \nonumber \\
&=& (\varPi_x \varphi, \varPi_x \varphi)_+^{1/2}  
\langle   1_{Q_- + x} A \nabla  \varphi,  \nabla \varphi \rangle^{1/2}.
\end{eqnarray}
Together with \eqref{eq:lemma2a} this yields the assertion since 
 $(\varPi_x \varphi, \varphi)_+ =   (\varPi_x \varphi, \varPi_x \varphi)_+$.  
 
\noindent (ii): Since $\sum_{x \in  \T_N} 1_{Q_- + x}(y) = l^d$ for all $y \in \T_N$ the inequalities follow
by summing (i) over $x \in  \T_N$. If $(\Tscr \varphi, \varphi)_+ = 0$ then $(\varPi_x \varphi, \varphi)_+ = 0$
for all $x \in \T_N$ and thus $\varPi_x \varphi = 0$ and $P_x \varphi = \varphi$. 
Lemma \ref{lemma:projection} implies that there exist constants $c_x$ such that
$(\Ascr \varphi)(y) = c_x$ for all $y \in Q +x$. Since $l \geq 3$ the cubes $Q + x$ and $Q +(x + {\rm e}_i)$
overlap and this yields $c_x = c_{x+{\rm e}_i}$ for all $i = 1, \ldots, d$. Thus $c_x$ is independent of $x$. 
Since $\Ascr \varphi \in \cbX_N$ this implies $c=0$. Hence $\Ascr \varphi = 0$ and therefore $\varphi = 0$. \\
Now suppose that $(\Tscr \varphi, \varphi)_+ = (\varphi, \varphi)_+$. This implies that
for all $x \in \T_N$ we have $(\varPi_x \varphi, \varphi)_+ = \langle  1_{Q_-+x} A \nabla \varphi, \nabla \varphi\rangle$.
We claim that the last identity implies that $\nabla \varphi(x) = 0$. Indeed, if $1_{Q_-+x} \nabla \varphi = 0$ we
are done. Otherwise the identity can only hold if the inequality in \eqref{eq:lemma2b} is an identity. 
In particular we must have $\nabla \varPi_x \varphi = \lambda  1_{Q_- +x} \nabla \varphi$ and $\lambda = 1$.
Now $\varPi_x \varphi$ vanishes outside $Q +x $ and in particular at the points $x$ and $x + {\rm e}_i$.
Thus $\nabla \varPi_x \varphi(x) = 0$ and hence $\nabla \varphi(x) = 0$. It follows that $\varphi$ is constant on 
$\T_N$ and hence $\varphi = 0$ since $\varphi$ has mean zero. 

\noindent (iii): It follows from (ii) that $(\varphi, \psi)_{*} := (\Tscr \varphi, \psi)_+$ defines a scalar product on $\cbH_+$.
Thus the Cauchy Schwarz inequality and (ii) yield
\begin{equation}
(\Tscr \varphi, \psi)_+  \leq (\Tscr \varphi, \varphi)_+^{1/2} (\Tscr \psi, \psi)_+^{1/2}
\leq  (\Tscr \varphi, \varphi)_+^{1/2} (\psi, \psi)_+^{1/2}.
\end{equation}
Taking $\psi = \Tscr \varphi$ we obtain the desired estimate.
\qed
\end{proofsect}

\noindent Consider  the operator ${\Tscr}^{\prime}\colon \cbH_- \to\cbH_- $ dual with respect to $\Tscr$ and defined by 
\begin{equation}
\langle {\Tscr}^{\prime}\varphi,\psi\rangle=\langle \varphi,\Tscr\psi\rangle,\quad \varphi\in\cbH_-, \psi\in\cbH_+.
\end{equation}
Notice that 
\begin{equation}
\label{E:T'T}
{\Tscr}^{\prime}=\Ascr{\Tscr}{\Ascr}^{-1}, 
\quad (\Tscr' \varphi, \psi)_-  = ( \varphi, \Tscr' \psi)_-, 
\quad \mbox{and   } (\Tscr' \varphi, \varphi)_- = (\Tscr \Ascr^{-1} \varphi,  \Ascr^{-1} \varphi)_+.
\end{equation}
Indeed, for any $ \varphi\in\cbH_+$, we have 
\begin{equation}
\langle {\Tscr}^{\prime}\Ascr\varphi,\psi\rangle=\langle \Ascr\varphi,\Tscr\psi\rangle=(\varphi,\Tscr\psi)_+=(\Tscr\varphi,\psi)_+=\langle\Ascr\Tscr\varphi,\psi\rangle,
\end{equation}
and this yields the first identity in \eqref{E:T'T}. Now 
\begin{equation}
(\Tscr' \varphi, \psi)_-  = \langle \Ascr^{-1}  \Ascr \Tscr \Ascr^{-1} \varphi, \psi \rangle
= \langle \Tscr \Ascr^{-1} \varphi, \Ascr \Ascr^{-1} \psi \rangle = (\Tscr \Ascr^{-1} \varphi, \Ascr^{-1} \psi)_+.
\end{equation}
Since the last expression is symmetric in $\varphi$ and $\psi$ we get the second identity in 
(\ref{E:T'T}) and taking $\psi= \varphi$ we obtain the third identity. 
%Given that ${\Tscr}^{\prime}$ comutes with translations in $\T_N$, there exists ${\Tcal}^{\prime}$ so that
%\begin{equation}
%({\Tscr}^{\prime}\varphi)(x)=\sum_{y\in\T_N}{\Tcal}^{\prime}(x-y)\varphi(y).
%\end{equation}
Similarly, we have  ${\varPi}_x^{\prime}=  \Ascr{\varPi_x}{\Ascr}^{-1}$ for  the dual of ${\varPi}_x$. Notice that
\begin{equation}
\label{E:Pi'=0}
{\varPi}_x^{\prime}\varphi=0 \quad \mbox{whenever    } \supp \varphi\cap (Q+x)=\emptyset. 
\end{equation}
Indeed, considering any test function $\psi\in \cbX_N $, we have 
$\langle {\varPi}_x^{\prime}\varphi,\psi\rangle=\langle\varphi, {\varPi}_x \psi\rangle=0$.
We also consider the  operator  
\begin{equation}
{\Rscr}:=\id-{\Tscr} \quad  \mbox{and its dual   } {\Rscr}^{\prime}=\id-{\Tscr}^{\prime}.
\end{equation}
% as well as the corresponding kernels ${\Rcal}=\delta-{\Tcal}$ and ${\Rcal}^{\prime}=\delta-{\Tcal}^{\prime}$ with $\delta(x-y)$ denoting the Kronecker %delta.
It follows from Lemma \ref{lem2}(ii)  and (\ref{E:T'T}) that
\begin{equation} \label{eq:positivityT'}
(\Tscr' \varphi, \varphi)_- > 0, \quad (\Rscr' \varphi, \varphi)_- > 0,
\quad (\Tscr' \varphi, \Tscr' \varphi)_- \leq (\Tscr' \varphi, \varphi)_-   
\  \text{ for all }\   \varphi \in \cbH_- \setminus \{0\}.
\end{equation}

We next discuss the locality properties of translation invariant  bilinear forms, operators and the corresponding kernels.
These properties would be obvious if we consider bilinear forms on $\cbV_N$ since then we can use the Dirac masses $\delta_x$ as
test functions. Dirac masses, however, do not belong to $\cbX_N$ and hence we need to use test functions with broader support
which makes the conclusion of Lemma \ref{lemma:locality} below nontrivial. We begin by recalling the relation between bilinear forms, 
operators and kernels. The translation operator $\tau_a, a\in\T_N $, is defined by $\tau_a \varphi(y) = \varphi(y-a)$, so that $\tau_a \delta_y = \delta_{a+y}$. Recall that $\cbM_N$ denotes the space 
of  all matrix-valued, $L^N$ periodic maps with zero mean.

\begin{lemma} \label{lemma:representation} 
Let $B$ be a translation invariant bilinear form on $\cbX_N$, i.e.,
\begin{equation}
B(\tau_a \varphi, \tau_a \psi) = B( \varphi, \psi) \  \text{ for all }\   \varphi, \psi \in \cbX_N, \  \text{ for all }\   a \in \T_N.
\end{equation}
Then the following assertions hold.
\begin{enumerate}
\item[(i)] There exists a unique linear operator $\Bscr\colon \cbX_N \to \cbX_N$ such that
\begin{equation}
\langle \Bscr \varphi, \psi \rangle = B( \varphi, \psi) \  \text{ for all }\   \varphi, \psi \in \cbX_N.
\end{equation}
Moreover $\Bscr$ is translation invariant, i.e.,
$\Bscr \tau_a = \tau_a \Bscr$.
\item[(ii)]
There exists a unique matrix-valued kernel $\Bcal \in  \cbM_N$ such that
\begin{equation}
(\Bscr \varphi)(x) = \sum_{ y \in \T_N} \Bcal(x-y) \varphi(y) \  \text{ for all }\   x \in \T_N,  \  \text{ for all }\   \varphi \in \cbX_N.
\end{equation}
Moreover for $\tilde{\Bcal}\colon \T_N \to \R^{m \times m}$ we have
\begin{equation}
(\Bscr \varphi)(x) = \sum_{ y \in \T_N} \tilde{\Bcal}(x-y) \varphi(y) \  \text{ for all }\   x \in \T_N  \  \text{ for all }\   \varphi \in \cbX_N.
\end{equation}
if and only if
\begin{equation}
\tilde{\Bcal} - \Bcal = C
\end{equation}
with a constant $m\times m$ matrix $C$.
\item[(iii)]
If $\Bcal' \in \cbX_N$ denotes the kernel of the dual operator $\Bscr'$ then
\begin{equation}
\Bcal'(z) = \Bcal(-z).
\end{equation}
\item[(iv)]
If $\Bscr_1$ and $\Bscr_2$ are translation invariant operators on $\cbX_N$ and $\Bscr_3 = \Bscr_1 \Bscr_2$ then $\Bscr_3$
is translation invariant and the corresponding kernels $\Bcal_i \in \cbM_N, i=1,2,3$, are related by discrete convolution, i.e., 
\begin{equation}
\Bcal_3(x) = (\Bcal_1 * \Bcal_2)(x) := \sum_{y \in \T_N}  \Bcal_1(x-y) \Bcal_2(y)\quad \mbox{ for all } x\in\T_N.
\end{equation}
\end{enumerate}
\end{lemma}

\begin{proofsect}{Proof}
We include the elementary proof for the convenience of the reader. \\
(i):  Existence and uniqueness of $\Bscr$ follows from the Riesz representation theorem. To prove translation invariance
let $\Dscr := \Bscr \tau_a - \tau_a \Bscr$. We have
\begin{equation}
\langle \Dscr \varphi, \tau_a \psi\rangle = \langle \Bscr \tau_a \varphi, \tau_a \psi \rangle 
 - \langle \tau_a \Bscr \varphi, \tau_a \psi \rangle.
 \end{equation}
Since $\tau_a$ is an isometry with respect to the scalar product $\langle \cdot, \cdot \rangle$ we get
\begin{equation}
\langle \Dscr \varphi, \tau_a \psi\rangle = B(\tau_a \varphi, \tau_a \psi)  - \langle \Bscr \varphi, \psi \rangle
= B(\tau_a \varphi, \tau_a \psi) - B( \varphi,  \psi)  = 0.
\end{equation}
This holds for all $\varphi, \psi \in \cbX_N$. Hence $\Dscr = 0$.  

\noindent (ii): To show the existence of $\Bcal$ note that $(\delta_0 - \frac{1}{L^{Nd}}) a\in \cbX_N$ with any $a \in \R^m$ and define
\begin{equation}
\Bcal(x) a:= \Bscr [a (\delta_0 - \tfrac{1}{L^{Nd}} )](x).
\end{equation}
Using that $\tau_y(\delta_0)=\delta_y$, we get
\begin{multline}
\Bcal(x-y) a= \Bscr [a (\delta_0 - \tfrac{1}{L^{Nd}} )](x-y)= \tau_y\bigl(\Bscr [a (\delta_0 - \tfrac{1}{L^{Nd}} )]\bigr)(x) =\\= \Bscr \tau_y[a (\delta_0 - \tfrac{1}{L^{Nd}} )](x)=\Bscr [a (\delta_y - \tfrac{1}{L^{Nd}} )](x).
\end{multline}
Observing, further, that  for any $\varphi \in \cbX_N$, we have
\begin{equation}
\varphi = \sum_{ y \in \T_N} \varphi(y)  (\delta_y - \tfrac{1}{L^{Nd}}),
\end{equation}
we get
\begin{equation}
(\Bscr \varphi)(x)= \sum_{y \in \T_N}     \Bscr  [\varphi(y)(\delta_y - \tfrac{1}{L^{Nd}} )](x)
= \sum_{y \in \T_N} \Bcal(x-y) \varphi(y).
\end{equation} 
This shows the existence of $\Bcal$. Suppose now that
$\Bscr \varphi(x) = \sum_{y\in\T_N}\tilde{\Bcal}(x-y) \varphi(y)$ for all $\varphi \in \cbX_N$ and set $C(z) = \tilde{\Bcal}(z) - \Bcal(z)$. 
The choice $\varphi = (\delta_{{\rm e}_i} - \delta_0)a$ with an arbitrary $a\in\R^m$ yields
\begin{equation} 
C(x-{\rm e}_i) = C(x) \  \text{ for all }\   x \in \T_N, \  \text{ for all }\   i = 1, \ldots, d.
\end{equation}
Thus $\tilde{\Bcal} - \Bcal = C$ with a constant $m\times m$ matrix $C$. If in addition $\tilde{\Bcal} \in \cbM_N$ then this implies
$\tilde{\Bcal} = \Bcal$. Conversely if $\tilde{\Bcal} = \Bcal + C$ then
$\tilde{\Bcal}$ and $\Bcal$ generate the same operator since $\sum_{ x \in \T_N} \varphi(x) = 0$ for
$\varphi \in \cbX_N$.

\noindent (iii):  We have
\begin{eqnarray}
\langle  \Bscr' \varphi, \psi \rangle & = & \langle \varphi, \Bscr \psi \rangle
= \sum_{x,y \in \T_N} \langle\varphi(x), \Bcal(x-y) \psi(y) \rangle_{\R^m}
=  \sum_{x,y \in \T_N} \langle\varphi(y), \Bcal(y-x) \psi(x)\rangle_{\R^m} \nonumber \\
&=&  \sum_{x,y \in \T_N}  \langle\Bcal(-[x-y])\varphi(y),  \psi(x)\rangle_{\R^m}.
\end{eqnarray}
Hence $(\Bscr' \varphi)(x) = \sum_{y \in \T_N}  \Bcal(-[x-y]) \varphi(y)$ and the uniqueness
result in (ii) implies that $\Bcal'(z) = \Bcal(-z)$.

\noindent (iv):  One easily verifies that $\Bcal_1 * \Bcal_2 \in \cbM_N$ and that
$(\Bscr_1 \Bscr_2 \varphi)(x) = \sum_{y \in \T_N} \Bcal_1 * \Bcal_2(x-y) \varphi(y)$. 
Thus the assertion follows from the uniqueness result in (ii). 
\qed
\end{proofsect}

For two sets $M_1, M_2 \subset \T_N$  we define
\begin{equation}
\dist_\infty(M_1,M_2) := \min \{ \rho_\infty(x,y)\colon x \in M_1, y \in M_2 \}.
\end{equation}

\begin{lemma} \label{lemma:locality}
Let $B$ be a translation invariant bilinear form on $\cbX_N$ and let $\Bscr$ and $\Bcal \in \cbM_N$ be the associated operator and
the associated  kernel, respectively. Let $n$  be an integer and suppose
that $L^N > 2n + 3$.
Then the following three statements are equivalent.
\begin{enumerate}
\item[(i)]\ \ 
$B(\varphi, \psi) = 0 \quad \mbox{whenever   } \dist_\infty(\supp \varphi, \supp \psi) > n .$
\item[(ii)]\ \ 
There exists  an $m\times m$ matrix $C$ such that 
$\Bcal(z) =C \quad  \mbox{whenever   } \rho_\infty(z, 0) > n.$
\item[(iii)]\ \ 
$ \supp \Bscr \varphi \subset \supp \varphi + \{-n,  \ldots, n\}^d  \  \text{ for all }\    \varphi \in \cbX_N.$
\end{enumerate}
\end{lemma}

\begin{proofsect}{Proof} The implication     (ii) $\Longrightarrow$ (iii) is  easy.
Set $\tilde{\Bcal}(z) = \Bcal(z) - C$. Then $\tilde{\Bcal}(z) = 0$ if $\rho_\infty(z) > n$ with $\rho_\infty(z)=\rho_\infty(z,0) $ and
by Lemma \ref{lemma:representation}(ii) we have
\begin{equation}
(\Bscr \varphi)(x) = \sum_{y \in \T_N} \tilde{\Bcal}(x-y) \varphi(y).
\end{equation}
If $x \not \in  \supp \varphi + \{-n,  \ldots, n\}^d$ then either $y \not \in \supp \varphi$ or
$ y\in\supp \varphi $ and $\rho_\infty(x-y,0) > n$. In either case $\Bscr \varphi(x) = 0$.  

The implication (iii) $\Longrightarrow$ (i) is also  easy.
Suppose  that $\dist_\infty(\supp \varphi, \supp \psi) > n$.
Then (iii) implies that
$\dist_\infty(\supp \Bscr \varphi, \supp \psi) > 0$, i.e, $\Bscr \varphi$ and $\psi$ have
disjoint support.  Thus $B(\varphi, \psi) = \langle \Bscr \varphi, \psi \rangle = 0$.

To prove the  implication  (i) $\Longrightarrow$  (ii), consider the torus $\T_N$ with the fundamental domain $\L_N= \{ - \frac{L^N-1}{2}, \ldots,  \frac{L^N-1}{2}  \}^d$ and set
\begin{eqnarray}
M \,&:= &  \left\{ -n, \ldots, n \right\}^d,\\
M_-&:= &  \left\{   -n, \ldots, n+1 \right\}^d, \text{ and the closure of }  M,\\
\overline{M} \,&=& \left\{  -(n+1), -n, \ldots, n+1 \right\}^d.
\end{eqnarray}
Note that by the assumption $ L^N > 2n +3$  the set $\T_N\setminus \overline{M} $ is nonempty.

\noindent We first show that for all $i,j \in \{1, \ldots, d\}$ we have
\begin{equation}  \label{eq:localC}
\nabla_i \nabla^*_j \Bcal = 0  \quad \mbox{in   }    \T_N\setminus \overline{M} .
\end{equation}
To see this let $\xi \in  \T_N\setminus \overline{M} $ and consider
\begin{equation}
\psi =( \delta_{\xi + {\rm e}_i} - \delta_\xi)a,  \quad \varphi = (\delta_{{\rm e}_j} - \delta_0)b \text{ with } a,b\in\R^m.
\end{equation}
Since $|{\rm e}_i-{\rm e}_j|_\infty \leq 1$ we have
\begin{equation}
\dist_\infty(\supp \varphi, \supp \psi) \geq \rho_\infty(0, \xi) - 1 \geq n +1.
\end{equation}
Hence
\begin{equation}
0 =  B( \varphi, \psi) =  \langle \bigl(\Bcal(\xi + {\rm e}_i - {\rm e}_j) - \Bcal(\xi + {\rm e}_i)
- (\Bcal(\xi  - {\rm e}_j) - \Bcal(\xi))\bigr) a,b\rangle= \langle \nabla_i \nabla^*_j \Bcal(\xi)a,b\rangle
\end{equation}
for every $a,b\in\R^m$ and thus 
\begin{equation}
 \nabla_i \nabla^*_j \Bcal(\xi)=0.
\end{equation}
Next we show that for $j \in \{1, \ldots, d\}$ there exist a matrix $C_j$ such that
\begin{equation}  \label{eq:localD}
\nabla_j^* \Bcal = C_j  \quad \mbox{in  } \T_N\setminus \overline{M}.
\end{equation}
Fix $j$ and set $f = \nabla_j^* \Bcal$. 
Using the shorthand $I=\{-n-1, \dots, n+1\}$, let
\begin{equation}
x_1 \in \{ - \tfrac{L^N-1}{2}, \ldots,  \tfrac{L^N -1}{2} \} \setminus I,
\quad x' := (x_2, \ldots, x_d)\in  \{ - \tfrac{L^N-1}{2}, \ldots,  \tfrac{L^N -1}{2}\}^{d-1}.
\end{equation}
Then $x=(x_1, x') \in  \T_N\setminus \overline{M} $ and hence for $i \neq 1$
we have $(\nabla_i f)(x_1, x') = 0$.  Thus there exists a matrix-valued function 
$g_1$ on $  \{ - \tfrac{L^N-1}{2}, \ldots,  \tfrac{L^N -1}{2} \} \setminus I$
such that
\begin{equation}
f(x) = g_1(x_1) \quad \mbox{if   } x_1 \not \in I.
\end{equation}
In the same manner, there exists a  function $g_2$ such that $f(x) = g_2(x_2)$ if $x_2 \not \in I$. This eventually implies
(\ref{eq:localD}).

Further,
\begin{equation}
\nabla_j^* \Bcal(-n-1, x') - \nabla_j^* \Bcal (-n-2, x') = (\nabla_1 \nabla_j^* \Bcal)(-n-2, x') = 0.
\end{equation}
Hence we also have $\nabla^*_j \Bcal(x) = C_j$ if $x_1 = -n-1$. Arguing similarly for the other components of $x$ we get
\begin{equation} 
 \label{eq:localE}
\nabla_j^* \Bcal = C_j \quad \mbox{in   }   \T_N \setminus  M_-.
\end{equation}

\noindent We now show that $C_j = 0$. Assume without loss of generality $j \neq 1$. For $x_1 \not \in (I\setminus\{-n-1\})$ we have
\begin{equation}
\nabla_j^* \Bcal(x_1, x') = C_j \  \text{ for all }\   x' \in  \{ - \tfrac{L^N-1}{2}, \ldots,  \tfrac{L^N -1}{2}  \}^{d-1}  . 
\end{equation}
On the other hand $\sum_{x' \in \{ - \frac{L^N-1}{2}, \ldots,  \frac{L^N -1}{2}\}^{d-1}}  \nabla_j^* \Bcal(x_1, x')= 0$
since $\Bcal$ is periodic and $j \neq 1$. Thus $C_j = 0$. 

Arguing as in the derivation of (\ref{eq:localD}) we conclude from (\ref{eq:localE})  and the fact that $C_j = 0$
that there exists a matrix $C$
such that 
\begin{equation}
\Bcal = C \quad \mbox{in   }  \T_N \setminus M_-.
\end{equation}
 In addition we have
 \begin{equation}
\Bcal(n+1,x') - \Bcal(n+2, x') = (\nabla_1^* \Bcal)(n+2, x') = 0.
\end{equation}
Hence $\Bcal(x) = C$ for $x_1 = n+1$. Arguing similarly for the other components of $x$ 
we get $\Bcal = C$ in $ \T_N \setminus M$. This finishes the proof of Lemma~\ref{lemma:locality} (ii).
\qed
\end{proofsect}

\begin{lemma}      \label{lemma:localityT}
Suppose that $\dist_\infty(\supp \varphi, \supp \psi) > l-1$. Then
\begin{equation}
\langle \Tscr \varphi, \psi \rangle = 0, \quad 
\langle \Tscr' \varphi, \psi \rangle = 0, \quad
\langle \Rscr \varphi, \psi \rangle = 0, \quad
\langle \Rscr' \varphi, \psi \rangle = 0.  
\end{equation}
\end{lemma}

\begin{proofsect}{Proof}
It suffices to prove the first identity. The second follows by exchanging $\varphi$ and $\psi$ and the 
third and fourth follow since $\Rscr = \id - \Tscr$ and $\Rscr' = \id -\Tscr'$. 
By Lemma \ref{L:PixPiy} we have
\begin{equation}
\varPi_x \varphi = 0  \quad \mbox{if   }  \supp \varphi \cap (\overline Q + x) = \emptyset
\end{equation}
and it follows from the definition of $\varPi_x$ that $\supp \varPi_x \varphi \subset Q +x$.
Assume $ \langle \Tscr \varphi, \psi \rangle \neq 0$. Then there exist $x \in \T_N$ such that
$ \langle \varPi_x \varphi, \psi \rangle \neq 0$. Thus  $ \supp \psi \cap (Q +x) \neq \emptyset$
and $\supp \varphi \cap (\overline Q + x) \neq \emptyset$. Therefore there exist  
$\xi\in \overline Q$ and $\zeta\in Q$  such that $x+\xi \in \supp \varphi$, $x+\zeta
\in \supp \psi$.
Thus 
\begin{equation}
x+\xi - (x+\zeta) = \xi - \zeta \in \left\{ -(l-1), \ldots, l-1 \right\}^d.
\end{equation}
Hence $\dist_\infty(\supp \varphi, \supp \psi) \leq l-1$.
\qed 
\end{proofsect}

 Consider now the inverse $\Cscr =\Ascr^{-1}$.
 (For time being, we omit the reference to the matrix $A$ in the notation for $\Cscr$.
We will reinstate it  in Section~\ref{sec-mainproofanalytic}, where we explicitly discuss the smoothness with respect to $A$.)
The main step toward  the decomposition, is to subtract a positive definite operator from $\Cscr $ in such a way that the remnant is  positive definite
 and of finite range.
We define 
\begin{equation}
\Cscr_1:=\Cscr- \Rscr \Cscr{\Rscr}^{\prime}, \  \text{ which yields} \  \Ccal_1=\Ccal- \Rcal* \Ccal*{\Rcal}^{\prime}.
\end{equation}

\begin{prop}\label{P:FRDprop1}
Both $\Cscr_1$ and $\Rscr \Cscr{\Rscr}^{\prime}$ are positive definite and  $\Cscr_1$ has finite range, i.e.,
\begin{equation}  \label{eq:finiterangeCscr}
\langle \Cscr_1 \varphi, \psi \rangle = 0 \quad \mbox{if   } \dist_\infty(\supp \varphi, \supp \psi) > 2l - 3.
\end{equation}
In particular,  there exists an $m\times m$ matrix $C$ such that
\begin{equation}  \label{eq:finiterangeCcal}
\Ccal_1(z)=C \text{ if } \  \rho_\infty(z,0)>2l - 3. 
\end{equation}
\end{prop}

\begin{proofsect}{Proof}
For any $\varphi,\psi \in\cbX_N $, we use  \eqref{E:Gdef}  to get
\begin{equation}
\langle\Rscr \Cscr{\Rscr}^{\prime}\varphi,\varphi\rangle=({\Rscr}^{\prime}\varphi,{\Rscr}^{\prime}\varphi)_-\ge 0.
\end{equation}
If $\Rscr' \varphi = 0$ then \eqref{eq:positivityT'} implies that $\varphi = 0$. Thus $\Rscr \Cscr \Rscr'$ is positive definite. Furthermore, 
\begin{equation}
\label{E:C_1}
\begin{aligned}
\langle\Cscr_1\varphi,\psi\rangle&=\langle\Cscr\varphi,\psi\rangle-
\langle\Cscr{\Rscr}^{\prime}\varphi, {\Rscr}^{\prime}\psi\rangle\\&=(\varphi, \psi)_--({\Rscr}^{\prime} \varphi, {\Rscr}^{\prime} \psi)_-=
({\Tscr}^{\prime}\varphi,  \psi)_-+(\varphi, {\Tscr}^{\prime} \psi)_--({\Tscr}^{\prime} \varphi, {\Tscr}^{\prime} \psi)_-.
\end{aligned}
\end{equation}
Thus \eqref{eq:positivityT'} implies that $\Cscr_1$ is positive definite. 

To evaluate the range of the quadratic form  $\langle \Cscr_1 \varphi, \psi \rangle$, we inspect the terms on the right hand side of \eqref{E:C_1}. For the first (and similarly the second) term, we have
\begin{equation}
({\Tscr}^{\prime}\varphi,\psi)_-=\frac{1}{l^d}\sum_{x\in  \T_N}({\varPi}_x^{\prime}\varphi,\psi)_-=\frac{1}{l^d}\sum_{x\in \T_N}({\varPi}_x^{\prime}\varphi,{\varPi}_x^{\prime} \psi)_-.
\end{equation}
In view of \eqref{E:Pi'=0},  a term in the sum vanishes at $ x $ except when the supports of $ \varphi $ and $ \psi $ both intersect $ Q+x $. Therefore, the scalar product is zero whenever the distance of the supports is strictly greater than $ l-1 $.  The second term of the bilinear form 
$ G_1(\varphi, \psi) :=  \langle \Cscr_1 \varphi, \psi \rangle$ is the double sum
\begin{equation}
( {\Tscr}^{\prime} \varphi, {\Tscr}^{\prime} \psi)_-=\frac{1}{l^d}\sum_{y\in \T_N}\frac{1}{l^d}\sum_{x\in \T_N}({\varPi}_y^{\prime}\varphi,{\varPi}_x^{\prime} \psi)_-.
\end{equation}
By Lemma~\ref{L:PixPiy} we have 
 ${\varPi}_x^{\prime}{\varPi}_y^{\prime}=\Ascr{\varPi}_x{\varPi}_y\Ascr^{-1}=0$ whenever
$(Q_-+x)\cap(Q_-+y) =\emptyset$, i.e., if $\rho_\infty(x,y) > l-1$. Hence the double sum only contains a non-zero contribution
if there exist $x$ and $y$ such that  $\rho_\infty(x,y) \leq l-1$,   $\supp \varphi \cap Q +x \neq \emptyset$, and
 $\supp \psi \cap Q +y \neq \emptyset$. Hence there must exist $\xi, \zeta \in Q$ such that
 $x +  \xi \in \supp \varphi$ and $y + {\zeta} \in \supp \psi$.
 Hence 
 \begin{equation}
\dist_\infty(\supp \varphi, \supp \psi) \leq \rho_\infty(x+   \xi-(y +  \zeta), 0)
\leq \rho_\infty(x-y, 0) + \rho_\infty(  \xi-\zeta, 0) \leq l-1 + l-2 \leq 2l- 3.
\end{equation}
This proves \eqref{eq:finiterangeCscr},  and \eqref{eq:finiterangeCcal} 
follows from Lemma \ref{lemma:locality}.
\qed
\end{proofsect}

We construct a finite range decomposition by an iterated application of Proposition~\ref{P:FRDprop1}. 
Let $L\ge 16$ and consider 
%% be divisible by $8$ and consider
\begin{equation} 
 Q_j=\{1,\ldots, l_j - 1\}^d \quad \mbox{with  }  l_j = \lfloor \tfrac18 L^j\rfloor + 1  \quad \mbox{for   }  j=1,\ldots, N.
 \end{equation}
Here $\lfloor a \rfloor$ denotes the integer part of $a$, the largest integer not greater than $a$. 
 In particular we have
 \begin{equation}   \label{eq:estimateslj}
         \tfrac18   L^j      <  \,    l_j  \leq  \tfrac18  L^j + 1.
 \end{equation} 
We define  ${\Tscr}_j,{\Tscr}_j^\prime $, and $ {\Rscr}_j^\prime $  as before with $ Q $ replaced by $ Q_j $ and set 
\begin{equation}
\label{E:C_k}
\Cscr_k : =(\Rscr_{1}\dots \Rscr_{k-1} )\Cscr(\Rscr_{k-1}^\prime\dots \Rscr_{1}^\prime)
- (\Rscr_{1}\dots \Rscr_{k-1} \Rscr_k)\Cscr(\Rscr_k^\prime  \Rscr_{k-1}^\prime  \dots \Rscr_{1}^\prime), \ k=1,\dots, N,
\end{equation}
and
\begin{equation}
\label{E:C_N+1}
\Cscr_{N+1}  :=(\Rscr_{1} \dots \Rscr_{N-1}\dots \Rscr_N)\Cscr(\Rscr_N^\prime  \Rscr_{N-1}^\prime  \dots \Rscr_{1}^\prime).
\end{equation}
With these definitions, we  show that the sequence $\{\Cscr_k\}_{k=1,\dots,N+1}$ yields a finite range decomposition.

\begin{prop}\label{P:FRDprop2}  Suppose that $L \geq 16$.    %%, $d \geq 2$.   Should we mention this ???
Then the 
operators $\Cscr_k$   %% $k=1,\dots,N+1$,  
satisfy 
\begin{enumerate}
 \item[(i)]  $ \Cscr=\sum_{k=1}^{N+1}\Cscr_k .$
\smallskip
\item[(ii)] $\Cscr_k$ is positive definite for $k = 1, \dots, N+1$.
\item[(iii)]  For $k = 1, \dots, N$ the range of $\Cscr_k$ is bounded by $ \tfrac{1}{2}L^k $, i.e., 
\begin{equation}  
\langle \Cscr_k \varphi, \psi \rangle = 0 \quad \mbox{if   } \dist_\infty(\supp \varphi, \supp \psi) >  \tfrac12 L^k
\end{equation}
and there exist  $m\times m$ matrices  $C_k$  such that
\begin{equation}  \label{eq:finiterangeCcalk}
\Ccal_k(z)=C_k \ \text{ if } \  \rho_\infty(z,0)> \tfrac12 L^k.
\end{equation}
\end{enumerate} 
\end{prop}

\begin{remark} \label{rem:smallrange}\hfill

\noindent
(i) Let $\delta \in (0,1/2)$. Then we can obtain the sharper conclusion 
 $\Ccal_k(z)=C_k$  if  $\rho(z,0)> \delta L^k$, provided that $L$ is large enough and we choose
 the integers   $l_j$  sufficiently  small, e.g. we may take  $l_j = \lfloor\delta L^j/ 4 \rfloor + 1$ if $L \geq 8/\delta$. 
 Of course the regularity estimates \eqref{regularityboundfield}  then depend on $\delta$ and degenerate for $\delta \to 0$.\\
(ii) The restriction $L \geq 16$ can be removed. If $ 6 \leq L \leq 15$ we can take
$l_1 = 3$ and for $j \geq 2$ define $l_j$ as before. One can easily check that in this case we still have
$-1 + 2 \sum_{j=1}^k (l_j - 1) \leq L^k/2$ and this yields the desired assertion (see the proof below). 
If $3 \leq L \leq 5$ one can skip the first few renormalization steps.
Formally one can take $ l_1 = l_2 = 2$ and define $l_j$ as before for $j \geq 3$. 
Then $\Tscr_1 = \Tscr_2 = 0$, $\Rscr_1 = \Rscr_2 = \id$, $\Cscr_1 = \Cscr_2 = 0$ and 
$-1 + 2 \sum_{j=3}^k (l_j -1) \leq L^k/2$.
\end{remark}

\begin{proofsect}{Proof}
Assertion (i) follows directly from the definition. 
To prove (ii), set
\begin{equation}
\varphi_k := \Rscr'_{k-1} \dots \Rscr'_1 \varphi,  \quad \psi_k := \Rscr'_{k-1} \dots \Rscr'_1 \psi, \quad k=1,\ldots, N+1.
\end{equation}
Inductive application of \eqref{eq:positivityT'} shows that $\varphi_k = 0$ implies $\varphi = 0$.
Now, directly from definitions, $\langle \Cscr_{N+1} \varphi, \varphi \rangle = (\varphi_{N+1}, \varphi_{N+1})_-$. Thus 
$\Cscr_{N+1}$ is positive definite. 
For $k =1, \dots, N$ we have
\begin{equation}
\langle \Cscr_k \varphi, \varphi \rangle = \langle (\Cscr - \Rscr_k \Cscr \Rscr_k') \varphi_k, \varphi_k \rangle.
\end{equation}
Hence by Proposition \ref{P:FRDprop1} we get $\langle \Cscr_k \varphi, \varphi \rangle \geq 0$ with equality only holding
if $\varphi_k = 0$, which implies $\varphi = 0$. Thus $\Cscr_k$ is positive definite.

\noindent (iii): In view of the equation  $\langle \Cscr_k \varphi, \psi \rangle = \langle (\Cscr - \Rscr_k \Cscr \Rscr_k') \varphi_k, \psi_k \rangle$,
Proposition \ref{P:FRDprop1} implies that
\begin{equation}
\langle \Cscr_k \varphi, \psi \rangle = 0  \quad \mbox{if   }   \dist_\infty(\supp \varphi_k, \supp \psi_k) > 2 l_k  -3.
\end{equation}
Iterative application of Lemma \ref{lemma:localityT} and Lemma \ref{lemma:locality} yields
\begin{equation}
\supp \varphi_k \subset \supp \varphi + \{-n_k, \dots, n_k\}^d, \quad
\supp \psi_k \subset \supp \psi+ \{-n_k, \dots, n_k\}^d,\quad n_k = \sum_{j=1}^{k-1}  (l_j - 1).
\end{equation}
Thus 
\begin{equation}
\langle \Cscr_k \varphi, \psi \rangle = 0  \quad \mbox{if   }   \dist_\infty(\supp \varphi_k, \supp \psi_k) > - 1 + 2 \sum_{j=1}^k (l_j -1).
\end{equation}
Now since $l_j -1 \leq \frac18 L^j$ and $\sum_{n=0}^\infty L^{-n} \leq 2$ we get 
$2 \sum_{j=1}^k (l_j -1) \leq \frac12 L^k$. This finishes the proof.
\qed 
\end{proofsect}

%%%%%%%%%%%%%
%%%%%%%%%%%%

\section{Estimates for fixed $A$}    \label{sec-mainproofestimates}

To prove the regularity bounds of Theorem~\ref{THM:FRD} and Theorem~\ref{THM:FRDfamily} we derive estimates for the  Fourier multipliers
of the relevant operators. To this end,  we first extend
the space   $\cbV_N$ to the set  $\cbV_N=\bigl(\C^m\bigr)^{L^{Nd}}$ of complex-valued vectors  with the subspace $\cbX_N$ defined, again, as the subset of functions $\varphi\in \cbV_N$ with vanishing sum, $\sum_{x\in\T_N}\varphi(x)=0$.
Various scalar products and norms are extended to complex-valued functions in the usual way,
$\langle \varphi, \psi\rangle :=  \sum_{x \in \mathbb T_N} \langle \varphi(x), \psi^*(y)\rangle_{\C^m}$,
$\varphi,\psi\in \cbV_N$,
 with $\psi^*(x)$ denoting the complex conjugate of $\psi(x)$. 
 
To introduce the discrete Fourier transform, consider the set of (scalar) functions
 $f_p(x)=\e^{i\langle p,x\rangle}, p\in \widehat{\T}_N$ , labelled by the dual torus
\begin{equation}
\widehat{\T}_N=\bigl\{p=(p_1,\dots, p_d)\colon p_j\in \bigl\{\tfrac{(-L^N+1)\pi}{ L^N}, \tfrac{(-L^N+3)\pi}{L^N},\ldots, 0,\dots, \tfrac{(L^N-1)\pi}{L^N}\bigr\}, 
j=1,\dots, d \bigr\}.
\end{equation}
The family  of functions $\bigl\{ L^{-\frac{Nd}2} f_p \bigr\}_{p \in \widehat{\T}_N}$
forms an orthonormal basis of $\C^{L^{Nd}}$.
For any $\psi \in \cbV_N$, we can define the Fourier transform component-wise,
\begin{equation} 
\label{eq:Fourier}
\widehat \psi(p) := \sum_{x \in \mathbb T_N} f_p(-x) \psi(x)   \ \text{ for   }\  p \in \widehat{\T}_N .
\end{equation}
For $\psi\in \cbX_N$, we get $\widehat \psi(0)= \sum_{x\in\T_n}\psi(x)=0$ with the inverse 
\begin{equation} 
\label{E:invFour}
\psi(x) =  L^{-Nd} \sum_{p \in \widehat{\T}_N \setminus \{0\} } f_p(x)  \widehat \psi(p)
\end{equation}
and 
\begin{equation}
\label{E:Planch}
\langle \varphi, \psi \rangle =  L^{-Nd}\sum_{p \in \widehat{\T}_N \setminus \{0\} } \langle \widehat \varphi(p), \widehat \psi(p) \rangle_{\C^m}.
\end{equation}
In the same way (component-wise) we get the Fourier transform for a matrix-valued kernel $\Kcal \in \cbM_N$,
\begin{equation} 
\label{E:MFourier}
\widehat \Kcal(p) := \sum_{x \in \mathbb T_N} f_p(-x) \Kcal(x)  .
\end{equation}

For a translation invariant operator $\Kscr\colon \cbX_N \to \cbX_N$ with the  kernel $\Kcal \in \cbM_N$, we get
\begin{equation}  
\label{eq:multiplier}
\widehat{\Kscr \psi}(p) = \widehat \Kcal(p) \widehat \psi(p).
\end{equation}
Indeed,
\begin{eqnarray}
\widehat{\Kscr \psi}(p)
&=&  \sum_{x \in \mathbb T_N} \sum_{y \in \mathbb T_N} \Kcal(x-y) \psi(y) f_p(-x)
= \sum_{x \in \mathbb T_N}\sum_{y \in \mathbb T_N} \Kcal(x-y) \psi(y) f_p(-(x-y))f_p(-y)=  \nonumber \\
&=&\sum_{z \in \mathbb T_N} \Kcal(z)  f_p(-z)  \sum_{y \in \mathbb T_N} \psi(y)  f_p(-y) 
=  \widehat \Kcal(p) \widehat\psi(p).
\end{eqnarray}
Henceforth, we call $\widehat \Kcal$  the Fourier multiplier of $\Kscr$.  
Applying the equality \eqref{eq:multiplier} with $\psi=a f_p$, $p\neq 0$,  we get
\begin{equation}
\label{eq:Kamultiplier}
\Kscr a f_p= \widehat{\Kcal}(p) a f_p.
\end{equation}
Indeed, taking into account that $\widehat{f_p}(p^\prime)=L^{Nd}\delta_{p,p^\prime}$,
we have $ \widehat{\Kscr a f_p}(p^\prime)= \widehat{\Kcal}(p^\prime) \widehat{af_p}(p^\prime)=\\ L^{Nd}\delta_{p,p^\prime}\widehat{\Kcal}(p^\prime)a = L^{Nd}\delta_{p,p^\prime}\widehat{\Kcal}(p)a$ and thus
\begin{equation} 
\Kscr a f_p(x) =  L^{-Nd} \sum_{p^\prime \in \widehat{\T}_N \setminus \{0\} } f_{p^\prime}(x)  \widehat{\Kscr a f_p}(p^\prime)=
\sum_{p^\prime \in \widehat{\T}_N \setminus \{0\} } f_{p^\prime}(x)\delta_{p,p^\prime}\widehat{\Kcal}(p)a=\widehat{\Kcal}(p) a f_p.
\end{equation}
Notice also that, by Lemma \ref{lemma:representation}, the kernel of a  product of two translation invariant operators is
given by the discrete convolution of the kernels and thus
\begin{equation} 
 \label{eq:product}
\widehat{\Kcal_1 \ast \Kcal_2}(p) = \widehat \Kcal_1(p) \widehat \Kcal_2(p).
\end{equation}

Now, we will study the Fourier multipliers of the operators $\Ascr =\nabla^* A\nabla$
as well as the operators $\Tscr$ and $\Rscr$ introduced in the previous section. 
 Given that $\nabla_j f_p = q_j(p) f_p$ with $q_j(p) = {\rm e}^{i p_j} -1$ and
$\nabla^*_j f_p = q^*_j(p) f_p$, $j=1,\dots,d$, 
we have
\begin{equation}  
\label{eq:Apax}
(\Ascr a f_p)_r
 = \sum_{j,k,s}  q^*_j(p)  A_{r,j;s,k} a_s q_k(p)  f_p\  \text{ for any } a\in\R^m,
\end{equation}
where $A_{r,j;s,k}$ are matrix elements of $A$, $(A(a\otimes q))_{r,j}=\sum_{s=1}^m\sum_{k=1}^d A_{r,j;s,k} a_s q_k$.
In view of \eqref{eq:Kamultiplier}, we get, for the Fourier multiplier $\widehat \Acal(p)$ of the operator $\Ascr$, the expession 
\begin{equation}  
\label{eq:Apa}
(\widehat \Acal(p) a)_r
 = \sum_{j,s,k}    q^*_j(p)  A_{r,j;s,k} a_s q_k(p)\  \text{ for any } a\in\R^m.
\end{equation}
Alternatively, we can express this in terms of corresponding quadratic form,
\begin{equation}  
\label{eq:Apvariational}
\langle \widehat \Acal(p) a, b  \rangle_{\C^m} = \langle A (a \otimes q(p)), b \otimes q(p)\rangle_{\C^{m\times d}}\  \text{ for any }
a,b\in\R^m.
\end{equation}

It follows that  the multiplier $\widehat \Acal(p)$ is Hermitian and positive definite. More precisely,  for any  $p\in \widehat{\T}_N \setminus \{0\}$,
we have
\begin{equation}    \label{eq:boundsAp}
\|\widehat \Acal(p)\| \leq   \|A\|  \, |p|^2 \ \text{ and } \  \| \widehat \Acal(p)^{-1}\| \leq \frac{\pi^2}{4 c_0 |p|^2}.
\end{equation}
Indeed, the first bound follows directly from \eqref{eq:Apvariational}
and the definition  of  the operator norm  $\|A\|$ of the linear map $A$, 
\begin{equation} 
\|A\|:=  \max\{   |A F| \colon F\in \C^{m\times d}, |F| \le1 \}.
\end{equation}
We also took into account that $|q(p)|^2\le  |p|^2$ as follows from the upper bound in the estimate
  $ \frac{4}{\pi^2} t^{2} \leq  |{\rm e}^{i t} - 1|^{2} \leq t^{2}$ 
valid  for all $t \in [-\pi, \pi]$. To get the second inequality, we use the lower bound \eqref{E:AFF} as well as the lower bound above, to get 
the estimate
$\langle \widehat \Acal(p) a, a \rangle_{\C^m}  \geq  \frac{4}{\pi^2}  c_0  |p|^2  |a|^2$
for any $p \in \hat \T_N$.

For the Fourier multiplier of $\Tscr$ defined in \eqref{E:T}, we first recall that the translation operator $\tau_x$ is defined by $(\tau_x \psi)(y) = \psi(y-x)$. 
Hence we have  $\varPi_x  \psi = \tau_x \varPi_0 (\tau_{-x} \psi)$. Note also that $\tau_{-x} f_p = {\rm e}^{i \langle x, p\rangle} f_p$.
Writing $\varPi$ as a shorthand for $\varPi_0$, we  get
\begin{equation}
\label{E:Tp}
\widehat \Tcal(p) a = l^{-d} \widehat{\varPi(a f_p)}(p)
\end{equation}
for any $a \in \C^m$. Indeed, applying $\Tscr$ to $a f_p$, we get
\begin{eqnarray}
l^d \Tscr (a f_p) (y) 
&=&  \sum_{x \in\T_N}  \varPi_x (a f_p) (y)  =  \sum_{x \in\T_N} \varPi (\tau_{-x} a f_p) (y -x) 
\nonumber \\
&=&   \sum_{x \in\T_N} \varPi ({\rm e}^{i \langle p, x\rangle} a f_p) (y-x)
=  \sum_{x \in\T_N} \varPi(a f_p)(y-x)   \, \,   {\rm e}^{i\langle p, x - y \rangle} {\rm e}^{i \langle p,  y \rangle} 
\nonumber \\
&=&  \sum_{z \in\T_N} \varPi(a f_p)(z)     {\rm e}^{- i\langle p, z \rangle}  \, {\rm e}^{i \langle p,  y \rangle} 
=   \widehat{\varPi(a f_p)}(p)    f_p(y).
\end{eqnarray}
Thus, $ \Tscr (a f_p)= l^{-d} \widehat{\varPi(a f_p)}(p)    f_p$ implying the claim 
by comparing with \eqref{eq:Kamultiplier}.

We now use the symmetry and boundedness of $\Tscr$, with respect to the  scalar product $(\cdot, \cdot)_+$, to  deduce the corresponding properties for $\widehat \Tcal(p)$. 
According to \eqref{eq:Kamultiplier}, we have
\begin{equation}
\label{E:<ATf_p>}
(\Tscr (a f_p), b f_p)_+ 
= \langle \Ascr \Tscr (a f_p), b f_p \rangle= L^{Nd}\langle \widehat \Acal(p) \widehat \Tcal(p) a , b  \rangle_{\C^m}
\end{equation}
for any $a,b \in \C^m$.
Combining this  with the fact that  $\Tscr$ is Hermitian with respect to  $(\cdot, \cdot)_+$ and with Lemma \ref{lem2}(ii),
 we infer that
\begin{equation}  \label{eq:propertiesAT}
\langle \widehat \Acal(p) \widehat \Tcal(p) a , b  \rangle_{\C^m} = \langle a, \widehat \Acal(p) \widehat \Tcal(p) b  \rangle_{\C^m} \text{ and }
0 \leq \langle \widehat \Acal(p) \widehat \Tcal(p) a , a  \rangle_{\C^m}  \leq \langle \widehat \Acal(p)  a , a  \rangle_{\C^m}
\end{equation}
for all$a,b \in \C^m$.
 Since $\widehat \Acal(p)$ is Hermitian and positive definite, it has a unique Hermitian positive definite square root
 $\widehat \Acal(p)^{1/2}$  with inverse $\widehat \Acal(p)^{-1/2}$. 
Applying \eqref{eq:boundsAp}, we get
 \begin{equation}  
 \label{eq:boundsA1/2p}
 \| \widehat \Acal(p)^{1/2}\| \leq  \|A\|^{1/2} |p|\  \text{ and }\  \| \widehat \Acal(p)^{-1/2}\| \leq  
\frac{\pi}{2 \sqrt{c_0}}  \frac1{|p|}.
\end{equation}

Setting, finally
 \begin{equation}
\widetilde \Tcal(p) := \widehat \Acal(p)^{1/2} \widehat \Tcal(p) \widehat \Acal(p)^{-1/2} = \widehat \Acal(p)^{-1/2} (\widehat \Acal(p)  \widehat \Tcal(p)) \widehat \Acal(p)^{-1/2}
\end{equation}
and
\begin{equation}
\widetilde \Rcal(p) := \widehat \Acal(p)^{1/2} \widehat \Rcal(p) \widehat \Acal(p)^{-1/2} = \1 - \widetilde\Tcal(p),
\end{equation}
we get the following bounds.
 
\begin{lemma}\label{Fouriermultipliers} 
The operators $\widetilde \Tcal(p)$ and $\widetilde \Rcal(p)$ are Hermitian (with respect to the standard scalar product on $\C^m$) and satisfy,for any  $p\in \widehat{\T}_N \setminus \{0\}$, the following bounds:
\begin{enumerate}
\item[(i)] there is  a constant $c<\infty $ (which depends only on $\|A\|$ and $c_0$, and the dimension $d$) such that  
\begin{equation} \label{eq:highpassT}
  \|  \1 - \widetilde \Rcal(p)\| =  \norm{\widetilde \Tcal(p)}  \leq  \min(1, c(|p| l)^4),
\end{equation}

\item[(ii)] there is a constant $c<\infty$ (which depends only on $\norm{A}$, $c_0$, and the dimension $d$) such that 
\begin{equation} \label{eq:lowpassR}
 \|\widetilde \Rcal(p)\| \leq  \min \bigl(1, \tfrac{c}{l} \bigl(\tfrac{1}{|p|} + 1\bigr)\bigr),
 \end{equation}
\end{enumerate} 
\end{lemma}

These estimates show that $\Tscr$ suppresses low frequencies, while $\Rscr$ suppresses high frequencies, reflecting the idea
that $P_x= \id - \varPi_x$ is a (locally) smoothing operator (cf. Remark \ref{rem:lowpasshighpass}).

\begin{proofsect}{Proof}
From \eqref{eq:propertiesAT}, we get
\begin{equation}  \label{eq:propertiesTtilde}
\langle \widetilde \Tcal(p) a , b  \rangle_{\C^m} = \langle a,  \widetilde \Tcal(p) b  \rangle_{\C^m} \ \text{ and }\
0 \leq \langle \widetilde \Tcal(p) a , a  \rangle_{\C^m}  \leq \langle   a , a  \rangle_{\C^m}
\end{equation}
for any $a,b \in \C^m$.
The operators $\widetilde \Tcal(p)$ and $\widetilde \Rcal(p)$ are thus Hermitian and 
$0 \leq \widetilde \Tcal(p)  \leq \1$  and, equivalently,  $ 0 \leq \widetilde \Rcal(p)  \leq \1$. This implies that
\begin{equation} \label{eq:easyboundTR}
\norm{\widetilde \Tcal(p)} \leq 1, \quad \norm{\widetilde \Rcal(p)} \leq 1. 
\end{equation}

\noindent (i): In view of (\ref{eq:easyboundTR}) we may assume that $|p| l \leq 1$. We first estimate the norm of  the Hermitian matrix
$\widehat \Acal(p) \widehat \Tcal(p)$. 
First, we show that
\begin{equation}   
 \label{eq:AT1}
l^d \langle \widehat \Acal(p) \widehat \Tcal(p) a, a \rangle_{\C^m} = \| \varPi (a f_p) \|_+^2
\end{equation}
To see this, we start from the right hand side,
\begin{multline}
\| \varPi (a f_p) \|_+^2=(\varPi  (a f_p), a f_p)_+  =\langle \varPi (a f_p), \Ascr ( a f_p )\rangle=
L^{-Nd}\langle \widehat{\varPi (a f_p)},  \widehat{\Ascr  (a f_p)} \rangle=
\\=L^{-Nd}\sum_{p^\prime\in\widehat\T_N\setminus\{0\}}\langle \widehat{\varPi (a f_p)}(p^\prime),  \widehat{\Acal}(p)a\widehat{f_p} (p^\prime)\rangle_{\C^m}
=\langle  \widehat{\varPi (a f_p)}(p) , \widehat \Acal(p) a  \rangle_{\C^m}=l^d \langle \widehat \Acal(p) \widehat \Tcal(p) a, a \rangle_{\C^m}.
\end{multline}
Here,  we first used the fact that $\varPi$ is an orthogonal projection with respect to $(\cdot, \cdot)_+$, passing to the second line we  used the equation $ \widehat{\Ascr  (a f_p)}(p^\prime)=\widehat{\Acal}(p)a\widehat{f_p} (p^\prime)$ obtained as the Fourier transform of \eqref{eq:Kamultiplier} for $\Ascr$, then 
the fact that $f_p(p^\prime)=L^{Nd} \delta_{p,p^\prime}$ and, finally, we applied the equation \eqref{E:Tp}.

Applying \eqref{eq:Kamultiplier} and using that $af_p\in \cbX_N$ (for $p\neq 0$) and thus also $\varPi(af_p)\in \cbX_N$, we get
\begin{equation}  
\label{eq:AT2}
 \norm{\varPi (a f_p)}_+^2   = \langle  \varPi (a f_p), \Ascr (a f_p) \rangle= 
 \langle  \varPi (a f_p), \widehat{\Acal}(p)a f_p \rangle =\langle  \varPi (a f_p), \widehat{\Acal}(p)a (f_p-1) \rangle .
 \end{equation}      
Further, given that $ \varPi (a f_p)$ is supported in $Q$ and $|f_p(z) - 1| \leq \sqrt{d}\, |p| l$
for $z \in Q$,we have $ \norm{\varPi (af_p)}_2=\norm{\varPi(af_p)}_{\ell_2(Q)} $
and  $\norm{(f_p-1)1_Q}_2= \norm{f_p-1}_{\ell_2(Q)}\leq \sqrt{d}\, \abs{p} l^{d/2+1} $. With the help of Schwarz inequality and this observation, we get
\begin{equation}
\label{E:paf}
 \norm{\varPi (a f_p)}_+ ^2 = \langle  \varPi (a f_p), \widehat{\Acal}(p)a (f_p-1)1_Q \rangle\leq
  \norm{\varPi (a f_p)}_{\ell_2(Q)}\norm{\widehat{\Acal}(p)} \abs{a} \sqrt{d}\, \abs{p} l^{d/2+1}.
\end{equation}
The Poincar\'e inequality (\cite{pach,che}) implies that 
\begin{equation}
 \norm{\varPi (a f_p)}_{\ell_2(Q)}  \leq \bar c l  \norm{\nabla \varPi (a f_p)}_{\ell_2(Q)} \leq \frac{\bar c}{c_0^{1/2}} l   \norm{\varPi (a f_p)}_+
\end{equation}
with a suitable constant $\bar c$. Combining this with \eqref{E:paf} and \eqref{eq:boundsAp}, we get
\begin{equation}
 \norm{\varPi (a f_p)}_+ \leq \sqrt{d}\,\bar c \frac{\|A\|}{c_0^{1/2}} |p|^3  l^{d/2+2} |a|.
\end{equation}
Given that $\widehat \Acal(p) \widehat \Tcal(p)$ is Hermitian, we get
\begin{equation}
\norm{\widehat \Acal(p) \widehat \Tcal(p)} = \max_{|a| \le 1}  \langle \widehat \Acal(p) \widehat \Tcal(p) a, a \rangle_{\C^m}  \leq d\,\bar c^2 (\|A\|^2/c_0) |p|^6 l^4 .
\end{equation}
The assertion follows using the second inequality in  \eqref{eq:boundsAp}.

\noindent (ii):  In view of (\ref{eq:easyboundTR}), it suffices to consider the case $|p| l \geq 1$. Again, we first
estimate $\norm{\widehat \Acal(p) \widehat \Rcal(p)}$.  Since $ \widehat \Rcal(p) = \1 - \widehat \Tcal(p)$ we get from (\ref{eq:AT1})
\begin{equation} 
\label{eq:AR1}
l^d \langle \widehat \Acal(p) \widehat \Rcal(p) a, a \rangle_{\C^m}  =   l^d  \langle \widehat \Acal(p) a, a\rangle -   (\varPi (a f_p), a f_p)_+  .
\end{equation}
Let $\omega$ be a cut-off function such that
\begin{equation}
\omega(z)  =1  \mbox{   if  } z \in   \overline{Q} \setminus Q, \quad 
\omega = 0 \mbox{  if  }  \dist(z, \overline{Q} \setminus Q) \geq    1 + \frac{1}{|p|}  , \quad 0 \leq \omega \leq 1\ \text{ and }  \
\quad |\nabla \omega | \leq \bar c |p|
\end{equation}
with a suitable constant $\bar c$.
By Lemma  \ref{lemma:projection}~(ii) we have
$\varPi(1-\omega)(a f_p) = (1- \omega) 1_Q  a f_p$. Hence
\begin{equation}
\label{E:Piaeta}
(\varPi  (a f_p),  a f_p)_+ = (\varPi (a  \omega  f_p), a f_p)_+ + (a (1 - \omega)  1_Q f_p, a f_p)_+
\end{equation}
and 
\begin{multline}
\label{E:etaQ}
(a (1 - \omega)  1_Q f_p, a f_p)_+   = \langle a (1 - \omega)  1_Q f_p, \Ascr (a f_p) \rangle =\\= \langle a (1 - \omega)  1_Q f_p, \widehat{\Acal}(p) a f_p \rangle  =  \sum_{z \in Q_-} (1 - \omega)   \, \, 
  \langle \widehat \Acal(p) a, a \rangle_{\C^m}.
  \end{multline}
Here, in the last step, we used that   $\sum_{z \in Q} (1 - \omega)=\sum_{z \in Q_-} (1 - \omega)$ since  $\omega = 1$ on $Q_-\setminus Q$. Using that $|Q_-| = l^d$, the equations \eqref{eq:AR1} and \eqref{E:Piaeta} with \eqref{E:etaQ} yield
\begin{equation}   \label{eq:AR1bis}
l^d \langle \widehat \Acal(p) \widehat \Rcal(p) a, a \rangle_{\C^m}  = - (\varPi(a  \omega  f_p), a f_p)_+  + \sum_{z \in Q_-} \omega   
\langle  \widehat \Acal(p) a, a \rangle_{\C^m}. 
\end{equation}
Given that $\omega$ is supported in a neighbourhood of order $ 1 + 1/|p|$ around the boundary $\overline{Q} \setminus Q$ of $Q$,  the last term is easily estimated
\begin{equation}  \label{eq:AR2}
\bigl| \sum_{z \in Q_-} \omega   
\langle\widehat  \Acal(p) a, a \rangle_{\C^m}    \bigr|
 \leq  4d l^{d-1}(1 +  \frac{1}{|p|}) \|\widehat \Acal(p)\| |a|^2 \leq 4d l^{d-1} (1 + \frac{1}{|p|})  \|A\| |p|^2  |a|^2.
\end{equation}
To bound the remaining term we introduce another cut-off function $\tilde \omega$  that satisfies
the following conditions,
\begin{equation}
\tilde \omega(z)  =1  \mbox{   if  }  \dist(z, \overline{Q} \setminus Q) \leq   2 +  \frac{2}{|p|}   , \quad 
\tilde \omega = 0 \mbox{  if  }  \dist(z, \overline{Q} \setminus Q) \geq   3 +  \frac{3}{|p|}  , \quad 0 \leq \tilde  \omega \leq 1 
\ \text{ and } \ |\nabla \tilde  \omega | \leq \bar c |p|.
\end{equation}
Then 
\begin{eqnarray}
(\varPi(a  \omega  f_p), a f_p)_+ 
&=&( a \omega f_p,   \varPi (a f_p))_+
=    ( a \omega f_p,   \varPi (a \tilde \omega f_p)_+    +    ( a \omega f_p,   (1-\tilde \omega) 1_Q  a f_p)_+ \nonumber \\
   & =&  ( a \omega f_p,   \varPi (a \tilde \omega f_p)_+  + \langle  \Ascr(a \omega f_p),   (1-\tilde \omega) 1_Q  a f_p \rangle
   \nonumber  \\
   &=& ( a \omega f_p,   \varPi (a \tilde \omega f_p))_+
\end{eqnarray}
since $1 - \tilde \omega$ and $ \Ascr(a \omega f_p)$ have disjoint support.
Thus 
\begin{equation}   \label{eq:AR3}
 | (\varPi  (a  \omega  f_p), a f_p)_+ | 
 \leq    \| a \omega f_p\|_+    \,  \| \varPi (a \tilde \omega f_p)\|_+
 \leq   \| a \omega f_p\|_+   \,  \| a \tilde \omega f_p\|_+  \leq \tfrac{c}4 \|A\| |p|^2    l^{d-1} (3 + \frac{3}{|p|}) |a|^2,  
\end{equation}
where we used that $\omega$ and $\tilde \omega$ are supported in a strip of size $3 + 3/ |p|$ around $\overline{Q} \setminus Q$, that $1/|p| \leq l$,
and that the gradients of $\omega$, $\tilde \omega$ and $f_p$ are bounded by $\bar c |p|$ and the constant $c$ in \eqref{eq:AR3} is  suitably chosen in  dependence on $\bar c$ and $d$.
The combination of (\ref{eq:AR1bis}),   (\ref{eq:AR2})  and  (\ref{eq:AR3}) now yields the estimate
\begin{equation}
\|\widehat \Acal(p) \widehat \Rcal(p) \| \leq  (\tfrac34c+4d)  \|A\| |p|^2  \frac{1}{l} (1 +  \frac{1}{|p|})  .
\end{equation}
In view of (\ref{eq:boundsAp}) this finishes the proof of (ii).
\qed
\end{proofsect}

As in the previous section assume that $L \geq 16$ and consider
\begin{equation} 
 Q_j=\{1,\ldots, l_j - 1\}^d \quad \mbox{with  }  l_j = \lfloor \tfrac18 L^j\rfloor + 1   \mbox{ for   }  j=1,\ldots, N.
 \end{equation}
 Also operators  ${\Tscr}_j,{\Tscr}_j^\prime $, and $ {\Rscr}_j^\prime $,  as well as
$\Cscr_k , \ k=1,\dots, N+1$, are defined as before (cf. \eqref{E:C_k} and \eqref{E:C_N+1}).

We define $\Ascr^{1/2}$ via the action of the corresponding Fourier symbol,  
$\widehat{\Ascr^{1/2} \varphi}(p) = \widehat \Acal^{1/2}(p) \widehat \varphi(p)$. Similarly we define
$\Ascr^{-1/2}$. 
Then the operators $\widetilde \Rscr_k := \Ascr^{1/2} \Rscr_k \Ascr^{-1/2}$ and
$\widetilde \Tscr_k := \Ascr^{1/2} \Tscr_k \Ascr^{-1/2}$
 are Hermitian.
For $k = 1, \ldots, N$ define
\begin{equation}
\widetilde \Mscr_k := \widetilde \Rscr_1 \ldots  \widetilde \Rscr_{k-1}   \widetilde \Rscr_k \ \text{ and }\  \widetilde \Mscr_0 := \id.
\end{equation}
Since $\tilde \Rscr_k = \id - \widetilde \Tscr_k$ and $\Cscr = \Ascr^{-1} = \Ascr^{-1/2} \Ascr^{-1/2}$ we have for $k = 1, \ldots N$
\begin{eqnarray}  \label{eq:Cproduct1}
\Cscr_k &=&  \Ascr^{-1/2} [ \widetilde \Mscr_{k-1} \widetilde \Mscr_{k-1}' - \widetilde \Mscr_{k}  \widetilde \Mscr_{k}']  \Ascr^{-1/2}   \nonumber \\
&=&  \Ascr^{-1/2} 
       [  (\widetilde \Tscr_k \widetilde \Mscr_{k-1}) \widetilde  \Mscr_{k-1}'  + \widetilde \Mscr_{k-1}  (\widetilde \Tscr_k \widetilde  \Mscr_{k-1})' 
   -  \widetilde \Tscr_k \widetilde \Mscr_{k-1} (\widetilde \Tscr_k \widetilde \Mscr_{k-1})' ]   \Ascr^{-1/2} 
\end{eqnarray}
and
\begin{equation}  \label{eq:Cproduct2}
\Cscr_{N+1} =   \Ascr^{-1/2}   \widetilde \Mscr_{N} \widetilde  \Mscr_{N}'     \Ascr^{-1/2} . 
\end{equation}

To estimate the corresponding kernels $\Ccal_k$ and their derivatives we use
formula \eqref{E:invFour} for the inverse Fourier transform. This yields
\begin{equation}
\sup_{x \in \T_N} \|\Ccal_k(x)\| \leq \frac{1}{L^{Nd}} \sum_{p \in \widehat \T_N\setminus \{0\} } \|\widehat \Ccal_k(p) \|
\end{equation}
and for any multiindex $ \alpha =(\alpha_1,\ldots,\alpha_d) $,
\begin{equation}  \label{eq:bound_inverse_Fourier}
\sup_{x \in \T_N} \|\nabla^\alpha \Ccal_k(x)\| \leq \frac{1}{L^{Nd}} \sum_{p \in \widehat \T_N\setminus \{0\} } \|\widehat \Ccal_k(p) q^\alpha \|
\leq \frac{1}{L^{Nd} }\sum_{p \in \widehat \T_N\setminus \{0\} } \norm{\widehat \Ccal_k(p)} |p|^{|\alpha|},
\end{equation}
where $q_j = {\rm e}^{i p_j} - 1$ and $q^\alpha = \prod_{j=1}^d q_j^{\alpha_j}$.

Finally, a bound on $\norm{\widehat \Ccal_k(p)}$ will be based on \eqref{eq:Cproduct1} (respectively, \eqref{eq:Cproduct2}) combined with  bounds on $\| \widetilde \Mcal_k(p) \|$ and $\| \widetilde \Tcal_{k+1}(p)  \widetilde  \Mcal_k(p) \|$. 
 
The estimate of the latter will depend on $|p|$.  Namely, slicing the dual torus into the  annuli
\begin{equation}
\label{E:Aj}
A_j:= \bigl\{ p \in \widehat \T_{N} \setminus \{0\} :  \pi L^{-j} \le |p| < \pi L^{-j +1}   \bigr\},\ j=1, ..., N,  
\end{equation}
with the complement
\begin{equation}
\label{E:A0}
A_0:=  \bigl\{ p \in \widehat \T_{N} \setminus \{0\} :  |p| \geq \pi  \bigr\},
\end{equation}
and defining, for any $c\ge 1$ and $j,k=0,1,\dots,N$, the step functions
\begin{equation}
\label{function1}
M_{k,c,L}(p) := 
\begin{cases} 1, &\mbox{if } p \in A_j,  j \geq k,\\
\frac{c^{k-j}}{L^{(k-j)(k-j+1)/2}}, &\mbox{if } p \in A_j,  j < k,\end{cases}
\end{equation}
and
\begin{equation}
\label{function2}
\widetilde M_{k,c,L}(p) :=   
\begin{cases} c   L^8  L^{4 (k-j)},  &\mbox{if } p \in A_j,  j \geq k,\\
\frac{c^{k-j}}{L^{(k-j)(k-j+1)/2}}, &\mbox{if } p \in A_j,  j < k,\end{cases}
\end{equation}
we have the following estimates.
\begin{lemma}  \label{lemma:estimateMk}
There exists a constant $c$ (depending only on $c_0$, $\|A\|$, and $d$) such that for  any  odd
$L \geq 16$ and any $N\in \N, N\ge 1$,
\begin{equation}
\| \widetilde \Mcal_k(p) \| \leq   
M_{k,c,L}(p)
\ \text{ for   }\  k= 0, \ldots, N,
\end{equation}
and 
\begin{equation}
\| \widetilde \Tcal_{k+1}(p)  \widetilde  \Mcal_k(p) \| \leq   
\widetilde M_{k,c,L}(p)  \mbox{ for    } k = 0, \ldots, N-1.
\end{equation}
\end{lemma} 

\begin{proofsect}{Proof}
Let $p \in A_j$. For  $k \leq j$ both bounds follow from the bounds $\| \widetilde \Rcal_n(p)  \| \leq 1$ for $ n=0,1,\ldots, k $ and 
$\| \widetilde \Tcal_{k+1}(p) \| \le   c ( L^{k+1} |p|)^4$.  Now, assume that $k > j$ and 
recall that  (after increasing the constant $c$ from \eqref{eq:lowpassR}  by a constant factor)
\begin{equation} \| \widetilde \Rcal_n(p) \| \leq \frac{c}{L^n} \Bigl(1 + \frac{1}{|p|}\Bigr) \leq \frac{c}{L^{n-j}}, \quad \mbox{for   }
n = j+1, \ldots N.
\end{equation}
Thus 
\begin{equation}  \prod_{n=1}^{k}    \| \widetilde \Rcal_n(p) \|  \leq        \prod_{n=j+1}^{k} 
 \| \widetilde \Rcal_n(p) \|  
\leq \prod_{n=j+1}^{k} \frac{c}{L^{n-j}} 
= \frac{c^{k-j}}{L^{(k-j)(k+1-j)/2}} .
\end{equation}
The first estimate for $k > j$ follows, with the second estimate implied since
$\|\widetilde \Tcal_{k+1}(p) \| \leq 1$.
\qed
\end{proofsect}

To combine these bounds for an  estimate on the Fourier multpliers  $\widehat \Ccal_k(p)$,
we use the following  Lemma.

\begin{lemma}  \label{lemma:sum_estimates}
Let  $n$ be a nonegative integer and  $ c \geq 1$.   Then there 
exists a constant $c'$  (depending on parameters $c$, $n$, and the dimension $d$) such
that with 
\begin{equation}
\label{E:eta}
\eta=\eta(n,d)= \max(\tfrac14 (d+n-1)^2, d+n+6)+2
\end{equation}
%%  Notice that for $d+n\ge 8$ it is $\eta= \tfrac14 (d+n-1)^2+2$, 
%%  while for $d+n\le 7$ it is $\eta= d+n+8$.
%%  In particular, for $d=3$ we have
%%  $\eta= \tfrac14 (n+2)^2+2$ for $n\ge 6$ while
%%  $\eta=n+11$ for $n\le 5$.
 and for all integers $L \geq 3$,  $N\ge 1$, and all $k = 1, \ldots, N+1 $, we have   
\begin{equation}
\label{E:MtildeM}
\frac{1}{L^{dN}} \sum_{p \in \widehat \T_N \setminus \{0\} }     M_{k-1,c,L}(p) \widetilde M_{k-1,c,L}(p) |p|^{n-2}     \leq   c'  L^\eta  L^{-(k-1)(d+n-2)},
\end{equation}
\begin{equation}
\label{E:tildeMtildeM}
\frac{1}{L^{dN}} \sum_{p \in \widehat \T_N \setminus \{0\} }      \widetilde M_{k-1,c,L}(p) \widetilde M_{k-1,c,L}(p) |p|^{n-2}     \leq   c c'  L^{\eta+8}  L^{-(k-1)(d+n-2)},
\end{equation}
\begin{equation}
\label{E:MM}
\frac{1}{L^{dN}} \sum_{p \in \widehat \T_N \setminus \{0\} }       M_{N,c,L}(p)   M_{N,c,L}(p) |p|^{n-2} \leq   c'  L^\eta  L^{-N}(d+n-2).
\end{equation}
\end{lemma}

\begin{proofsect}{Proof}
It suffices to prove the first bound for $k= 1, \ldots N+1$. 
The second and third bounds follow employing the inequalities $\widetilde M_{k-1,c,L} \leq c L^8 M_{k-1,c,L}$ and 
$M_{N,c,L} \leq \widetilde M_{N,c,L}$, respectively.

To prove the first estimate we split the sum into the sum of contributions over the annuli $A_j$. 
For  $p\in A_j$, we have
\begin{equation}  
\label{eq:bound_p}
|p|^{n-2}  \leq \pi^{n-2} d^{n/2} L^2 L^{(-j+1)(n-2)}.
\end{equation}
Indeed, for $j\neq0$, we get
\begin{equation}
|p|^{n-2} \leq L^{\max((2-n),0)} 
 L^{(-j+1)(n-2)} \pi^{n-2} 
\leq \pi^{n-2}  L^2 L^{(-j+1)(n-2)}.
\end{equation}
The expression $(2-n)$ in the term $\max ( (2-n), 0)$ stems from the fact that for $n=0$ and $n=1$, 
we actually employ the lower bound on $\abs{p}$ from  \eqref{E:Aj}.
For $j= 0$, we have $\pi\le\,|p| \leq \sqrt{d} |p|_\infty \leq \sqrt{d} \pi$ and thus  $ \abs{p}^{n-2}\le  \pi^{n-2} d^{ n/2}\le \pi^{n-2} d^{n/2} L^n$. 
The size of the annuli can be bounded as
\begin{equation} \label{eq:bound_volume}
\frac{|A_j|}{L^{dN}} \leq \pi^d L^{(-j+1)d}.
\end{equation}

As a result, for $j \ge k-1$,
\begin{multline}
 \frac{1}{L^{dN}} \sum_{p \in A_j}  M_{k-1,c,L}(p) \widetilde M_{k-1,c,L}(p)  |p|^{n-2}
\leq  c\pi^{n+d-2}d^{n/2} L^2 L^{(-j+1) (d+n-2)} L^8 L^{4 (k-1-j)} \le \\
 \leq c \pi^{n+d-2}d^{n/2} L^{8+d+n}  L^{-(k-1)(d+n-2)} L^{-(j-(k-1))(d+n+2)}. 
\end{multline}
and
\begin{equation}  \label{eq:highj}
\frac{1}{L^{dN}} \sum_{j=k-1}^{N} \sum_{p \in A_j}  M_{k-1,c,L}(p) \widetilde M_{k-1,c,L}(p)  |p|^{n-2}
\leq \tilde c  L^{8+d+n}  L^{-(k-1)(d+n-2)}
\end{equation}
with $\tilde c= 2c \pi^{n+d-2}d^{n/2}  $ since 
\begin{equation}
\sum_{j=k-1}^{N}L^{-(j-(k-1))(d+n+2)}=\sum_{j^\prime=0}^{N-k+1}L^{-j^\prime(d+n+2)}\le\frac1{1-L^{-(d+n+2)}}.
\end{equation}

Now consider $j  < k-1$. We get
\begin{multline}
\frac{1}{L^{dN}}\sum_{p \in A_j}  M_{k-1,c,L}(p)
\widetilde M_{k-1,c,L}(p)    |p|^{n-2}
\leq \pi^{n+d-2}d^{n/2} L^2  L^{(-j+1) (d+n-2)}  \frac{c^{2(k-1-j)}}{L^{(k-1-j)(k-j) }   }
\leq
 \\
\leq  \pi^{n+d-2}d^{n/2} L^2 L^{(-k+1) (d+n-2)}  L^{(k-j)(d+n-2)}    \frac{c^{2(k-1-j)}}{L^{(k-1-j)(k-j)} }  
\end{multline}
Setting $j' = k-1-j$ we get
\begin{equation}   \label{eq:lowj}
\frac{1}{L^{dN}}\sum _{j=0}^{k-2} \sum_{p \in A_j}  M_{k-1,c,L}(p) \widetilde M_{k-1,c,L}(p)  |p|^{n-2}
\leq  \tilde c L^2   L^{-(k-1)(d+n-2)}     \sum_{j'=1}^{k-1}  
  \frac{c^{2j'}}{L^{j'(j'+1) - (d+n-2)(j'+1) } } .
\end{equation} 

Consider the integer $\bar \ell = \lfloor \frac{\log(2c^2)}{\log 3} \rfloor +1$ and split the sum above into terms with $j'\le \bar j$ and the rest with $j'>\bar j$, where $\bar j= d+n-2 + \bar\ell$. We get,
\begin{multline}
\label{E:sumj^2}
\sum_{j'=1}^{k-2}    \frac{c^{2j'}}{L^{j'(j'+1) - (d+n-2)(j'+1) } }\le
c^{2\bar j} \sum_{j'=1}^{\bar j}   L^{(d+n-2-j')(j'+1) }+
   \sum_{j'=\bar j+1}^{\infty}    \frac{c^{2j'}}{L^{(j'+1)\bar \ell } }\le\\
\le   \bar j c^{2\bar j} L^{\frac14(d+n-1)^2}+\frac{(\frac12)^{\bar j +1}}{L^{\bar \ell}(1 -\frac12) }\le 
\bar j c^{2\bar j} L^{\frac14(d+n-1)^2}+1
\end{multline}
Here, in the first sum, we bounded $(d+n-2-j')(j'+1)$ (with maximum at $j'= \frac{d+n-3}2$) by $ \frac14(d+n-1)^2$
and, in the second sum, we took into account that $L\ge 3$ and thus $c^2 L^{-\bar \ell} \le \frac12$.
 
Combining \eqref{eq:highj} and \eqref{eq:lowj} with \eqref{E:sumj^2}, 
we get the sought bound  for \eqref{E:MtildeM} and \eqref{E:MM} with constants $c'=\tilde c(2+\bar j c^{2\bar j} )$ and $\eta=\max(\tfrac14 (d+n-1)^2+2, d+n+8) $. For \eqref{E:tildeMtildeM}, the constants  must be increased by adding 8 to $\eta$ and multiplying $c'$  by $c$.
\qed
\end{proofsect}

\begin{proofsect}{Proof of Theorem~\ref{THM:FRD}}
By (\ref{eq:Cproduct1}), Lemma \ref{lemma:estimateMk}, and  the bound \eqref{eq:boundsA1/2p} we have  $\| \widehat \Ccal_k(p) \| \leq 2  c M_{k-1,c,L}(p) \widetilde M_{k-1,c,L}(p)  |p|^{-2} 
+  c \widetilde M_{k-1,c,L}(p) \widetilde M_{k-1,c,L}(p)  |p|^{-2}  $.
Now, for $k=1, \ldots, N$, the desired bounds  follow from     Lemma  \ref{lemma:sum_estimates}   and (\ref{eq:bound_inverse_Fourier}). 
For $k=N+1$ we use (\ref{eq:Cproduct2}) to get $\| \widehat \Ccal_{N+1}(p) \| \leq   c M_{N,c,L}(p)^2   |p|^{-2}$ and the assertion follows again
from    Lemma  \ref{lemma:sum_estimates}  and (\ref{eq:bound_inverse_Fourier}). 
\qed
\end{proofsect}

\bigskip

\section{Analytic dependence on $A$ and proof of Theorem~\ref{THM:FRDfamily}}
\label{sec-mainproofanalytic}

We now study the dependence of the finite range decomposition on the map $A$ which appears in the operator $\Ascr = \nabla^* A \nabla$. 
To this end we will show that the operators $ \varPi_x$, $ \Tscr$ and $\Rscr$
can be locally extended to holomorphic functions of $A$  and the bounds derived previously for fixed $A$ can be extended to a small
complex ball. Then the Cauchy integral formula immediately yields bounds on all derivatives with respect to $A$. 
We do not claim that the extensions of $ \varPi_x$, $ \Tscr$, $\Rscr$ to complex $A$ yield a finite range decomposition for complex $A$
(indeed positivity is meaningless if $\Ascr$ is not Hermitian). The extension is merely a convenient tool to show that the relevant quantities
are real-analytic as  functions of real, symmetric, positive definite $A$.

Let $A$ be a linear map from $\C^{m \times d}$ to $\C^{m \times d}$ such that
\begin{equation} \label{eq:definitionA}
A = A_0 + A_1 
\end{equation} 
with $A_0$ and $A_1$ satisfying the following  conditions 
\begin{equation}  \label{eq:conditionA0}
\langle A_0 F, G\rangle_{\C^{m\times d}} =\langle F, A_0 G \rangle_{\C^{m\times d}}, \quad  \langle A_0 F, F\rangle_{\C^{m\times d}} \geq c_0  \abs{F}^2, \  \text{ for all }\   F, G \in \C^{m \times d}, \ \text{ and }
\end{equation}
\begin{equation} \label{eq:conditionA1}
\| A_1 \| \leq \frac{c_0}{2}.
\end{equation}
Here, $c_0 > 0$ is a fixed constant and, as before, 
 $\langle \cdot, \cdot\rangle_{\C^{m\times d}} $ and $\abs{\cdot}$ denote the standard scalar product and norm on $\C^{m\times d}$ and
$\|A_1\|$ is  the corresponding operator norm of $A_1$. 

Again, we consider  the operator
\begin{equation} \label{eq:complexAscr}
\Ascr := \nabla^* A \nabla,
\end{equation}
on $\cbX_N$, i.e., 
\begin{equation} 
(\Ascr \varphi)_r := \sum_{j = 1}^{d} \nabla_j^* (A \nabla \varphi)_{j,r}, \
\mbox{ where   } \   (\nabla \varphi)_{j, r} = \nabla_j \varphi_r .
\end{equation}
 With operator $\Ascr$ we associate  the sesquilinear form
\begin{equation} \label{eq:sesquilinearA}
(\varphi, \psi)_A := \langle  A \nabla \varphi, \nabla \psi \rangle 
\end{equation}
where $ \langle\cdot,\cdot\rangle $ is the $ \ell_2 $-scalar product on $ \cbX_N$, defining the adjoint $\Ascr^*$ by
\begin{equation}
\langle \Ascr \varphi, \psi \rangle = (\varphi, \psi)_A  = \langle \varphi, \Ascr^*  \psi \rangle,
\ \mbox{ with   } \ \Ascr^* = \nabla^* A^* \nabla,
\end{equation}
where $A^*$ is the adjoint of $A$. Note that for real, symmetric $A$ the form $(\cdot, \cdot)_A$ is 
a scalar product and agrees with $(\cdot, \cdot)_+$. 
In the following, we  use the previous notation $\cbH_+$ for the Hilbert space with the scalar product $(\cdot, \cdot)_{A_0}$
and define $\|\varphi\|_{A_0} := (\varphi, \varphi)_{A_0}^{1/2}$. 

Using $\Re z$ and $z^*$ to denote the real part and the complex conjugate of a complex number $z$, we summarize the main properties of the sesquilinear form  $(\cdot, \cdot)_A$.

\begin{lemma}   \label{lemma:formA}
Assume that an operator $A$ satisfies the conditions \eqref{eq:definitionA}, \eqref{eq:conditionA0}, and \eqref{eq:conditionA1}.

\noindent
Then the  sesquilinear form $(\cdot, \cdot)_A$ on $\cbX_N$ satisifes
\begin{equation}   \label{eq:formAcoercive}
\Re (\varphi, \varphi)_A \geq \tfrac12  \|\varphi \|_{A_0}^2,
\end{equation}
\begin{equation}  \label{eq:formAbounded}
|(\varphi, \psi)_A | \leq \tfrac32 \|\varphi \|_{A_0}  \|\psi \|_{A_0},
\end{equation}
\begin{equation} \label{eq:formAsymmetry}
(\psi, \varphi)_A = (\varphi, \psi)_{A^*}^*.
\end{equation}
\end{lemma}

\begin{proofsect}{Proof}
The first claim follows using the definition of the form $(\cdot, \cdot)_A$ and the lower bound
\begin{equation} \label{eq:matrixAcoercive}
\Re \langle A F, F\rangle_{\C^{m\times d}}  \geq  \langle A_0 F, F\rangle_{\C^{m\times d}} - \frac{c_0}{2} \abs{F}^2 \geq \frac12 \langle A_0 F, F\rangle_{\C^{m\times d}}
\end{equation}
implied by \eqref{eq:conditionA0} and \eqref{eq:conditionA1}.

Using \eqref{eq:conditionA1}, the Cauchy-Schwarz 
inequality for the scalar product  $\langle A_0 F, G\rangle_{\C^{m\times d}}$, and the bound from \eqref{eq:conditionA0},
we also get
\begin{equation} \label{eq:matrixAbounded}
\begin{aligned}
|\langle A F, G\rangle_{\C^{m\times d}}| 
&\leq  \langle A_0 F, G\rangle_{\C^{m\times d}} + \frac{c_0}{2} \abs{F} \abs{G}
\leq \langle A_0 F, F\rangle_{\C^{m\times d}}^{1/2} \langle A_0 G, G \rangle_{\C^{m\times d}}^{1/2}  +\\
& + \tfrac12 \langle A_0 F, F\rangle_{\C^{m\times d}}^{1/2}  \langle A_0 G, G \rangle_{\C^{m\times d}}^{1/2} 
\leq \tfrac32 \langle A_0 F, F\rangle_{\C^{m\times d}}^{1/2}  \langle A_0 G, G \rangle_{\C^{m\times d}}^{1/2}
\end{aligned}
\end{equation}
implying the second claim.

 The last identity
 follows from the relation
 $\langle AG, F\rangle_{\C^{m\times d}} = \langle G, A^* F\rangle_{\C^{m\times d}} = \langle A^*F, G\rangle^*_{\C^{\m\times d}}$.
 \qed
 \end{proofsect}
In view of the above Lemma, the complex version of the Lax-Milgram theorem can be used to ensure
the existence of of the bounded inverse operator  $\Cscr_{A}= \Ascr^{-1}$.

In the following, similarly as in the case of the Hilbert space $\cbH_+$, we use $\cbH_+(Q +x)$ to denote the corresponding Hilbert space (of  functions from $\cbX_N$ with support in $Q+x$) with the scalar product $(\cdot, \cdot)_{A_0}$.

Next, we define an extension of the operators $\varPi_x$ for a general complex $A$.

\begin{lemma} Assume that  $A$ satisfies \eqref{eq:definitionA}, \eqref{eq:conditionA0}, and \eqref{eq:conditionA1}.
Then, for each $\varphi \in \cbX_N$, there exists a unique  $ v \in \cbH_{+}(Q +x)$ such that
\begin{equation}  \label{eq:etaprojection}
(v, \psi)_A = (\varphi, \psi)_A  \  \text{ for all }\   \psi \in \cbH_{+}(Q +x).
\end{equation}
\end{lemma}

\begin{proofsect}{Proof}
The assertion  follows from Lemma \ref{lemma:formA} and the Lax-Milgram theorem. 
\qed
\end{proofsect}

\begin{lemma} \label{lemma:propertiesPiA}
Assume that  $A$ satisfies  \eqref{eq:definitionA}, \eqref{eq:conditionA0}, and \eqref{eq:conditionA1}. 
For any $ \varphi\in\cbX_N $, we set
\begin{equation} \label{def:PiA}
 \varPi_{A,x} \varphi := v, \quad \varPi_A := \varPi_{A,0},
\end{equation}
with $v \in \cbH_+(Q+x)$  defined by \eqref{eq:etaprojection}.
Using, as before,  $\tau_x$ to denote the translation by $x$, $1_Q$ for the characteristic function of a set $Q$, 
and $D$ for  the open unit disc $D = \{ w \in \C\colon |w| < 1 \}$,
we have
\begin{enumerate}\itemsep-10pt
\item[(i)]\ \ $\varPi_{A,x} \tau_x \varphi = \tau_x \varPi_A \varphi$,\\
\item[\rm(ii)]\ \  $\|\varPi_A \varphi \|_{A_0} \leq 3 \|\varphi  \|_{A_0}$, \\
\item[\rm(iii)]\ \  $\varPi_A \varphi = \varphi  \  \text{ for all }\   \varphi \in \cbH_+(Q),  \quad \varPi_A \varPi_A = \varPi_A$,\\
\item[\rm(iv)]\ \  $\varPi_A \varphi = \varphi  1_Q  \quad   \text { if   }  \varphi = 0 \mbox{  on  } \overline{Q} \setminus Q$, \\
\item[\rm(v)]\ \  $(\varPi_A \varphi, \psi)_A = (\varphi, \varPi_{A^*}\psi)_A$,  \\
\item[\rm(vi)]\ \   The map $z \mapsto \varPi_{A_0 + z A_1} \varphi $ is holomorphic for $z$ in the open unit disc $D$.
\end{enumerate}
%\begin{flalign}
%&\text{\rm(i)}    &&\varPi_{A,x} \tau_x \varphi = \tau_x \varPi_A \varphi,  &&\label{LPiA1}\\
%&\text{\rm(ii)}      &&\|\varPi_A \varphi \|_{A_0} \leq 3 \|\varphi  \|_{A_0},   &&\label{LPiA2}\\
%&\text{\rm(iii)}    &&\varPi_A \varphi = \varphi  \  \text{ for all }\   \varphi \in \cbH_+(Q),  \quad \varPi_A \varPi_A = \varPi_A,  &&\label{LPiA3}\\
%&\text{\rm(iv)} &&\varPi_A \varphi = \varphi  1_Q  \quad  \mbox{if   }  \varphi = 0 \mbox{  on  } \overline{Q} \setminus Q, &&\label{LPiA4}\\
%&\text{\rm(v)}  &&(\varPi_A \varphi, \psi)_A = (\varphi, \varPi_{A^*}\psi)_A,   &&\label{LPiA5}\\
%&\text{\rm(vi)} &&\text{The map } z \mapsto \varPi_{A_0 + z A_1} \varphi \text{ is holomorphic for $z$ in the open unit disc } D.  &&\label{LPiA6}
%\end{flalign}
\end{lemma}

\begin{proofsect}{Proof}\\
(i): Given that the shift $\tau_{-x}$ is an isometry with respect to $(\cdot, \cdot)_{A_0}$ and maps $\cbH_+(Q + x)$ onto $\cbH_+(Q)$,
we have  the identities
$ (\tau_x \varPi_A \varphi, \psi)_A = (\varPi_A \varphi, \tau_{-x} \psi)_A = (\varphi, \tau_{-x} \psi)_A =
(\tau_x \varphi, \psi)_A$ for all  
$\psi \in \cbH_+(Q+x)$. As $\tau_x \varPi_A \varphi \in \cbH_+(Q +x)$, this yields the assertion. \\
(ii): Taking $\psi = varPi_A\varphi$ in the definition (\ref{eq:etaprojection}) of $v=\varPi_A\varphi$ and using Lemma \ref{lemma:formA},
we get 
\begin{equation}
\tfrac12 \|\varPi_A \varphi\|^2_{A_0} \leq \Re (\varPi_A  \varphi, \varPi_A \varphi)_A
= \Re (\varphi, \varPi_A \varphi)_A \leq \tfrac32 \|\varphi\|_{A_0}  \|\varPi_A \varphi \|_{A_0}. 
\end{equation}   
(iii): The second assertion follows from the first. By definition, we have $(\varPi_A \varphi - \varphi, \psi)_A = 0$
for all $\psi \in \cbH_+(Q)$. In particular, by the assumption $\varphi \in \cbH_+(Q)$ we may take $\psi = \varPi_A  \varphi - \varphi$ inferring that
$\varPi_A  \varphi = \varphi$.  \\
(iv): Let $\tilde \varphi =  \varphi 1_Q$. By (iii) we have $\varPi_A \tilde \varphi = \tilde \varphi$. Moreover
$\varphi - \tilde \varphi$ vanishes in $\bar Q$. Thus $\nabla (\varphi - \tilde \varphi)$ vanishes in $Q_-$.
Hence $(\varphi - \tilde \varphi, \psi)_A = 0$ for all $\psi \in \cbH_+(Q)$ since $\nabla \psi$ is supported in $Q_-$.
Therefore $\varPi_A (\varphi - \tilde \varphi) = 0$
which yields the assertion.  \\
(v) Since $\varPi_{A^*} \psi  \in \cbH^+(Q)$ we have
\begin{equation}
( \varphi, \varPi_{A^*} \psi)_A =  (\varPi_A \varphi, \varPi_{A^*} \psi)_A
=  (\varPi_{A*} \psi, \varPi_A \varphi)_{A^*}^*
=  ( \psi, \varPi_A \varphi)_{A^*}^*
= (\varPi_A \varphi, \psi)_A,
\end{equation}
where we used the relation $(\varphi, \psi)_A = (\psi, \varphi)_{A^*}^*$ and the definition of $\varPi_{A^*}$. \\
(vi): This follows from the complex inverse function theorem. Fix $\varphi$ and consider the map  $R$ from $ D \times \cbH^+(Q)$
into the dual of $\cbH^+(Q)$ given by
\begin{equation}
R(z, v)(\psi)  =(v- \varphi, \psi)_{A_0 + z A_1} = \langle (A_0 + z A_1) (\nabla v - \nabla \varphi), \nabla \psi \rangle .
\end{equation}
Then $R$ is complex linear in $z$ and $v$ and hence complex differentiable. By the definition of 
$\varPi_A$ we have $R(z, v) = 0$ if and only if $v = \varPi_{A_0 +z A_1} \varphi$. Finally the derivative
of $R$ with respect to the second argument is given by the map $L_z$ from $\cbH^+(Q)$ into its dual
with $L_z (\dot v) (\psi) = (\dot v, \psi)_{A_0 + z A_1}$. By the Lax-Milgram theorem, $L_z$ is
invertible for $z \in D$. Hence the map $z \mapsto \varPi_{A_0 + z A_1} \varphi$ is complex differentiable in $z$. 
\qed
\end{proofsect}
Note that for a real symmetric $A$ the above definition of $\varPi_{A,x}$ agrees with the definition of $\varPi_x$ in 
Section \ref{sec-mainproof}. We define, as before,
\begin{equation}
\Tscr_A  := l^{-d}  \sum_{x \in \T_N}  \varPi_{A,x},   \quad  \Rscr_A = \id - \Tscr_A .
\end{equation}
Then  the following weaker version of Lemma \ref{lem2} holds.

\begin{lemma}  \label{lemma:boundTcomplex}
Assume that $A$ satisfies  \eqref{eq:definitionA}, \eqref{eq:conditionA0}, and \eqref{eq:conditionA1}. 
 Then
\begin{equation}
\| \Tscr_A \varphi \|_{A_0}  \leq 9  \|\varphi \|_{A_0} \  \text{ for all }\   \varphi \in \cbX_N.  
\end{equation}
\end{lemma}

\begin{proofsect}{Proof}
This is an adaptation of the argument  from  \cite{BT06} to the complex case. For the convenience, we include the details. 
We have
\begin{equation} \label{eq:boundTcomplex1}
l^{2d} \| \Tscr_A \varphi \|_{A_0}^2 \leq 2 \, l^{2d}\,  | (\Tscr_A \varphi, \Tscr_A \varphi)_A |
\leq 2 \sum_{x, y \in \T_N}  |(\varPi_{A,x} \varphi, \varPi_{A,y} \varphi)_A| .
\end{equation}
Set $T_x := \nabla \varPi_{A,x} \varphi$. Then $T_x$ vanishes outside $Q_- +x$ since $\varPi_{A,x} \varphi$ vanishes outside $Q+x$. 
Thus,  in view of \eqref{eq:sesquilinearA} and \eqref{eq:formAbounded}, we get, similarly as in \eqref{eq:matrixAbounded},
 \begin{equation}
\begin{aligned}
\bigl|(\varPi_{A,x}\varphi,\varPi_{A,y}\varphi)_A 
\bigr|=\bigl|\langle A T_x, T_y\rangle\bigr|&=\big|\langle  A \1_{Q_-+x}  T_x,  \1_{Q_-+y} T_y\rangle\big|
=\big|\langle  A \1_{Q_-+y} T_x,\1_{Q_-+x} T_y\rangle\big|\le \\
&\le \tfrac{3}{2}\langle  A_0 \1_{Q_-+y}T_x,\1_{Q_-+y}T_x\rangle^{1/2}  \langle A_0 \1_{Q_-+x} T_y,\1_{Q_-+x} T_y\rangle^{1/2}\le \\
&\le       \tfrac{3}{4}\langle  A_0 \1_{Q_-+y}T_x,\1_{Q_-+y}T_x\rangle  + \tfrac34  \langle A_0 \1_{Q_-+x} T_y,\1_{Q_-+x} T_y\rangle=\\     
&=\tfrac34    \langle A_0 \1_{Q_-+y}T_x,T_x\rangle+\tfrac{3}{4}\langle A_0\1_{Q_-+x}T_y,T_y\rangle.
\end{aligned}
\end{equation}
Now $\sum_{y \in \T_N} 1_{Q_ +y}$ is the constant function $l^d$ and thus
\begin{eqnarray*}
\sum_{x,y \in \T_N} \bigl|(\varPi_{A,x}\varphi,\varPi_{A,y}\varphi)_A \bigr|
&\leq&  \tfrac32 l^d \sum_{x \in \T_N}   \langle A_0 T_x,T_x\rangle
=  \tfrac32 l^d \sum_{x \in \T_N}   (\varPi_{A,x} \varphi,  \varPi_{A,x} \varphi)_{A_0} \leq \\
&\leq&    3 l^d \sum_{x \in \T_N} \Re  (\varPi_{A,x} \varphi,  \varPi_{A,x} \varphi)_A
=   3 l^d \sum_{x \in T_N} \Re  ( \varphi,  \varPi_{A,x} \varphi)_A = \\
&=&   3  l^{2d}\Re  (\varphi, \Tscr_A \varphi)_A         
\leq  \tfrac92  l^{2d}  \|\varphi\|_{A_0} \| \Tscr_A \varphi\|_{A_0} .
\end{eqnarray*}
 Combined with \eqref{eq:boundTcomplex1}, this yields the assertion. 
\qed
\end{proofsect}

Next, we bound the Fourier multipliers of operators  $\Tscr_A$ and $\Rscr_A$.
Using the relation $\varPi_{A,x} = \tau_x \varPi_A \tau_{-x}$ we get, as before,
\begin{equation}
\Tscr_A  (af_p) = l^{-d} \sum_{z \in \T_N} {\rm e}^{-i \langle p, z \rangle} \varPi_A ( a f_p)(z)\, \,   f_p
\end{equation}
and thus, the Fourier multiplier $\widehat \Tcal_A(p)$ is given by
\begin{equation}
\widehat \Tcal_A(p)   a  = l^{-d}\sum_{z \in \T_N} {\rm e}^{-i \langle p, z \rangle} \varPi_A ( a f_p)(z)  . 
\end{equation}
Also, the operator $ \Ascr = \nabla^* A \nabla$ satisfies again the equation
\begin{equation}
 \label{eq:Apvariational2}
\Ascr (a f_p) = (\widehat \Acal(p) a)f_p 
\end{equation}
with
\begin{equation}
 \label{eq:Apvariational2}
 \langle \widehat \Acal(p) a, b  \rangle_{\C^m} = (A (a \otimes q(p)), b \otimes q(p)\rangle_{\C^{m\times d}}, \ q(p)_j = {\rm e}^{i p_j} - 1 .
\end{equation}
 Hence,
\begin{equation}    \label{eq:boundsApcomplex}
  \|\widehat \Acal(p)\| \leq \tfrac32  \|A_0\|  \, |p|^2 \text{ and } \ \| \widehat \Acal(p)^{-1}\| \leq \frac{\pi^2}{2 c_0 |p|^2}. 
\end{equation}

\begin{lemma}
Assume that $A$ satisfies  \eqref{eq:definitionA}, \eqref{eq:conditionA0}, and \eqref{eq:conditionA1}.  
Then, there is a constant $c<\infty$ (depending only on  $\|A_0\|, c_0$, and $d$) such that, for all $p \in \widehat \T_N\setminus\{0\}$,
\begin{enumerate}
\item[(i)]\ \ 
$\|    \widehat \Acal_0(p)^{1/2} \widehat \Tcal_A(p) \widehat \Acal_0(p)^{-1/2} \|  \leq 9, \quad
\|    \widehat \Acal_0(p)^{1/2} \widehat \Rcal_A(p) \widehat \Acal_0(p)^{-1/2} \|  \leq 10.$
\item[(ii)]\ \ 
$\| \widehat \Tcal_A(p) \| \leq  c  \min\bigl(1, (|p| l)^4\bigr).$
\item[(iii)]\ \ 
$\| \widehat \Rcal_A(p) \| \leq  c \min\bigl(1, \frac{1}{l} (1 + \frac{1}{|p|})\bigr).$
\end{enumerate}
\end{lemma}

\begin{proofsect}{Proof} The proof is largely parallel to the proof of Lemma~\ref{Fouriermultipliers} for real symmetric $A$, but the constants are slightly worse since $(\cdot, \cdot)_A$ is no longer a scalar product.

\noindent (i): The second bound follows from the first since $\widehat \Rcal_A(p) = \1 - \widehat \Tcal_A(p)$. For the first estimate, we apply Lemma \ref{lemma:boundTcomplex}
with $\varphi = a f_p$, $a \in \C^m$. This yields
\begin{equation}
\langle \widehat \Acal_0(p) \widehat \Tcal_A(p) a, \widehat \Tcal_A(p) a \rangle_{\C^m} \leq 81  \langle \widehat \Acal_0(p) a, a \rangle_{\C^m}.
\end{equation}
Taking $a = \widehat \Acal_0(p)^{-1/2} b$, we deduce that
\begin{equation}
\langle \widehat \Acal_0(p)^{1/2} \widehat \Tcal_A(p)\widehat  \Acal_0(p)^{-1/2} b, \widehat \Acal_0(p)^{1/2} \widehat \Tcal_A(p) \widehat \Acal_0(p)^{-1/2} b \rangle_{\C^m} \leq
 81  \langle b, b \rangle_{\C^m} 
\end{equation}
and this finishes the proof of (i). 

\noindent (ii):  First, we assume that $|p|  l \geq 1$. It follows from (i) that $\| \widehat \Tcal_A(p) \| \leq 9 \| \widehat \Acal_0(p)^{-1/2}\|   \|\widehat \Acal_0(p)^{1/2}\|$.
By  (\ref{eq:boundsA1/2p}) we have  $ \|\widehat \Acal_0(p)^{1/2}\| \leq  \|A_0\|^{1/2} |p|$ and $ \|\widehat \Acal_0(p)^{-1/2}\|  \leq 
\frac{\pi}{2\sqrt{c_0}}\frac1{|p|}$, yielding (ii).

\noindent  Now, assume that $|p| l \leq 1$. We first estimate the norm of $\widehat \Acal(p) \widehat \Tcal_A(p)$. Note that
\begin{equation}    \label{eq:ATcomplex1}
l^d \langle \widehat \Acal(p) \widehat \Tcal_A(p) a, b \rangle_{\C^m} = \langle \widehat \Acal(p)  \sum_{z \in \T_N} \varPi_A (a f_p)(z) , b f_p(z) \rangle_{\C^m} 
= \langle  \varPi_A (a f_p), \Ascr^* (b f_p) \rangle  = (\varPi_A  (a f_p), b f_p)_A .
\end{equation}
Thus by Lemma \ref{lemma:propertiesPiA} (iii), (v), and \eqref{eq:formAbounded}, we get
\begin{equation}  \label{eq:ATcomplex2}
\begin{aligned}
l^d \abs{\langle \widehat \Acal(p)\widehat  \Tcal_A(p) a, b \rangle_{\C^m}} & = \abs{(\varPi_A \varPi_A (a f_p), b f_p)_A}
= \abs{( \varPi_A (a f_p), \varPi_{A^*} (b f_p) )_A }\\ &\leq  
\tfrac32  \| \varPi_A (a f_p) \|_{A_0}  \| \varPi_{A^*} (b f_p) \|_{A_0}.
\end{aligned}
\end{equation}
To estimate $ \| \varPi_A (a f_p) \|_{A_0} $ we use \eqref{eq:formAcoercive} and the fact that 
$(\varPi_A (\varphi), \varPi_A (\varphi) )_A=  (\varphi, \varPi_A (\varphi) )_A$ according to \eqref{eq:etaprojection} since  $\varPi_A (\varphi)\in \cbH_+(Q)$,  yielding
\begin{eqnarray}
\tfrac12   \| \varPi_A (a f_p) \|_{A_0} ^2
&\le&  | (\varPi_A (a f_p), \varPi_A (a f_p) )_A   |
= |  ( a f_p,   \varPi_A (a f_p) )_A   |
= |\langle  \Ascr (a f_p),   \varPi_A (a f_p) \rangle |      \nonumber  \\
&=& | \langle  (\widehat \Acal(p) a) f_p,  \varPi_A (a f_p) \rangle  |
= | \langle  (\widehat \Acal(p) a) (f_p - 1),  \varPi_A (a f_p) \rangle  |.
\end{eqnarray}
In the last step we used the fact that functions in $\cbX_N$ have average zero.
Now $ \varPi_A (a f_p)$ is supported in $Q$ and $|f_p(z) - 1| \leq \sqrt{d}\, |p| l$
for $z \in Q$. In combination with (\ref{eq:boundsApcomplex}) this yields
\begin{equation}
\tfrac12  \| \varPi_A (a f_p) \|_{A_0} ^2 
 \leq   \| (\widehat \Acal(p) a) (f_p - 1) 1_Q \|   \,   \|\varPi_A (a f_p) \|
 \leq \tfrac32 \|A_0\| |p|^2 \sqrt{d}\, |a|  (|p| l)   l^{d/2}    \|\varPi_A (a f_p) \| .
\end{equation}
The Poincar\'e inequality (\cite{pach,che}) implies that 
\begin{equation}
 \|\varPi_A (a f_p) \|_{\ell^2(Q)}  \leq \bar c l  \| \nabla \varPi_A (a f_p) \|_{\ell^2(Q)} \leq \frac{\bar c}{c_0^{1/2}} l   \|\varPi_A (a f_p) \| _{A_0}.
\end{equation}
Combining the inequalities above we get
\begin{equation}
 \| \varPi_A (a f_p) \|_{A_0}  \leq 3 \sqrt{d}\, \bar c \frac{\|A_0\|}{c_0^{1/2}} |p|^3 l^2 \, l^{d/2} |a|.
\end{equation}
The same estimate holds for $\varPi_{A*}$. 
Hence, from  (\ref{eq:ATcomplex2}),
\begin{equation}
l^d \abs{ \langle \widehat \Acal(p) \widehat \Tcal_A(p) a, b\rangle_{\C^m}} \leq  \tfrac{27}2 d \frac{\bar c^2\|A_0\|^2}{c_0} |p|^6 l^4 \, l^d |a| |b|.
\end{equation}
With the help of  \eqref{eq:boundsApcomplex}, this yields the claim for a suitable constant $c$.  \\
(iii): For $ |p| l \leq 1$  the estimate follows from (i).
Thus, we assume $ |p| l \geq 1$.  Again we first estimate $\widehat \Acal(p) \widehat \Rcal_A(p)$.
Since $ \widehat \Rcal_A(p) = \1 - \widehat \Tcal_A(p)$ we get from (\ref{eq:ATcomplex1})
\begin{equation} \label{eq:ARcomplex1}
l^d \langle \widehat \Acal(p) \widehat \Rcal_A(p) a, b \rangle_{\C^m}  =   l^d  \langle \widehat \Acal(p) a, b\rangle_{\C^m} -   (\varPi_A  (a f_p), b f_p)_A .
\end{equation}
Let $\omega$ be a cut-off function such that
\begin{equation}
\omega(z)  =1  \mbox{   if  } z \in  \overline{Q} \setminus Q, \quad 
\omega = 0 \mbox{  if  }  \dist(z, \overline{Q} \setminus Q) \geq    1 + \frac{1}{|p|}  , \quad 0 \leq \omega \leq 1,   
\quad |\nabla \omega | \leq \bar c |p|.
\end{equation}
By Lemma  \ref{lemma:propertiesPiA}~(iii), we have
$\varPi_A (1-\omega)(a f_p) = (1- \omega) 1_Q  a f_p$. Hence
\begin{equation}
(\varPi_A  (a f_p), b f_p)_A = (\varPi_A (a  \omega  f_p), b f_p)_A + (a (1 - \omega)  1_Q f_p, b f_p)_A
\end{equation}
and
\begin{eqnarray}
(a (1 - \omega)  1_Q f_p, b f_p)_A 
&=& \langle a (1 - \omega)  1_Q f_p, \Ascr^* (b f_p))_A 
= \langle a (1 - \omega)  1_Q f_p, (\widehat \Acal(p)^* b) f_p \rangle  \nonumber \\
&=& \sum_{z \in Q} (1 - \omega)   \langle \widehat \Acal(p) a, b \rangle_{\C^m} 
 = \sum_{z \in Q_-} (1 - \omega)  
  \langle \widehat \Acal(p) a, b \rangle_{\C^m}.
\end{eqnarray}
In the last step we used that $\omega = 1$ on $Q_-\setminus Q$. Since $|Q_-| = l^d$, this yields
\begin{equation}   \label{eq:ARcomplex1bis}
l^d \langle \widehat \Acal(p) \widehat \Rcal_A(p) a, b \rangle_{\C^m}  = - (\varPi_A (a  \omega  f_p), b f_p)_A  + \sum_{z \in Q_-} \omega   
\langle \Acal(p) a, b \rangle_{\C^m}. 
\end{equation}
Given that $\omega$ is supported in a neighbourhood of the order $1 + 1/|p|$ around $\overline Q\setminus Q$, the last term is bounded by
\begin{equation}  \label{eq:ARcomplex2}
\bigl| \sum_{z \in Q_-} \omega   
\langle\widehat  \Acal(p) a, b \rangle_{\C^m}    \bigr|
 \leq  4d l^{d-1} (1 + \frac{1}{|p|}) \|\widehat \Acal(p)\| |a| |b| \leq 6 d l^{d-1} (1 +  \frac{1}{|p|})  \|A_0\| |p|^2  |a| |b| .
\end{equation}
To estimate the remaining term we introduce again an additional cut-off function $\tilde \omega$ for which
\begin{equation}
\tilde \omega(z)  =1  \mbox{   if  }  \dist(z, \overline{Q} \setminus Q) \leq  2 +   \frac{2}{|p|}   , \quad 
\tilde \omega = 0 \mbox{  if  }  \dist(z, \overline{Q} \setminus Q) \geq    3 + \frac{3}{|p|}  , \quad 0 \leq \tilde  \omega \leq 1,   
\quad |\nabla \tilde  \omega | \leq \bar c |p|. 
\end{equation}
Then, taking into account that $1 - \tilde \omega$ and $ \Ascr(a \omega f_p)$ have disjoint support,
\begin{eqnarray}
(\varPi_A (a  \omega  f_p), b f_p)_A  
&=&( a \omega f_p,   \varPi_{A^*} (b f_p))_A 
=    ( a \omega f_p,   \varPi_{A^*} (b \tilde \omega f_p))_A  +    ( a \omega f_p,   (1-\tilde \omega) 1_Q  b f_p)_A \nonumber \\
   & =&  ( a \omega f_p,   \varPi_{A^*} (b \tilde \omega f_p))_A  + \langle  \Ascr(a \omega f_p),   (1-\tilde \omega) 1_Q  b f_p \rangle
   \nonumber  \\
   &=& ( a \omega f_p,   \varPi_{A^*} (b \tilde \omega f_p))_A.
\end{eqnarray}
Hence, 
\begin{eqnarray}   \label{eq:ARcomplex3}
 | (\varPi_A (a  \omega  f_p), b f_p)_A | 
 &\leq &\tfrac32  \| a \omega f_p\|_{A_0}    \, \|\varPi_{A^*} (b \tilde \omega f_p)\|_{A_0} 
 \leq \tfrac92   \| a \omega f_p\|_{A_0}  \,  \| b \tilde \omega f_p\|_{A_0}   \nonumber \\
 & \leq& C  \|A_0\| |p|^2    l^{d-1} (1 +  \frac{1}{|p|}) |a| |b|,  
\end{eqnarray}
where we used that $\omega$ and $\tilde \omega$ are supported in a strip of order $1 + 1/ |p|$ around $\bar Q \setminus Q$, that $1/|p| \leq l$
and that the gradients of $\omega$, $\tilde \omega$ and $f_p$ are bounded by $\bar c |p|$.
The combination of (\ref{eq:ARcomplex1bis}),   (\ref{eq:ARcomplex2})  and  (\ref{eq:ARcomplex3}) now yields the estimate
\begin{equation}
\|\widehat \Acal(p) \widehat \Rcal_A(p) \| \leq  C  \|A_0\| |p|^2  \tfrac{1}{l} (1 +  \tfrac{1}{|p|} ) .
\end{equation}
In view of (\ref{eq:boundsApcomplex}) this finishes the proof of (iii).
 \qed
\end{proofsect}

We now define and estimate the operators $\Cscr_{A,k}$.
Assuming again that $L \geq 16$, we use
\begin{equation} 
 Q_j=\{1,\ldots, l_j - 1\}^d \quad \mbox{with  }  l_j = \lfloor \tfrac18 L^j\rfloor + 1  \quad \mbox{for   }  j=1,\ldots, N,
 \end{equation}
to define  ${\Tscr}_{A,j},{\Tscr}_{A,j}^\prime $, $ {\Rscr}_{A,j} $, and $ {\Rscr}_{A,j}^\prime $,  in the same way as before. 
 Introducing  
\begin{equation}
 \Mscr_{A,k} := \Rscr_{A,1} \ldots  \Rscr_{A,k-1}   \Rscr_{A,k},  \ \text{ for } \  k = 1, \ldots, N, \ \text{ and }\  \Mscr_{A,0} := \id,
\end{equation}
we set

\begin{equation}
\label{E:CkA}
\Cscr_{A,k} : =\Mscr_{A,k-1}\Cscr_A\Mscr_{A,k-1}^\prime
- \Mscr_{A,k} \Cscr_{A}\Mscr_{A,k}^\prime, 
\end{equation}
for  $\ k=1,\dots, N$ and         
and 
\begin{equation}
\label{E:CN+1A}
\Cscr_{A,N+1}  :=\Mscr_{A,N}\Cscr_{A}\Mscr_{A,N}^\prime
\end{equation}
for $k = N + 1$.

Considering the same annuli $A_j$, $j=0,1,\dots, N$, introduced  in \eqref{E:Aj} and \eqref{E:A0},  as well as  the functions $M_{k,c,L}(p)$ and $\widetilde M_{k,c,L}(p)$ defined in
\eqref{function1} and   \eqref{function2}, we have the following bound.

\begin{lemma}  \label{lemma:estimateMkcomplex}
Assume that $A$ satisfies  \eqref{eq:definitionA}, \eqref{eq:conditionA0}, and \eqref{eq:conditionA1}. 
Then there exists a constant $c$ (depending only on $\|A_0\|, c_0$ and the dimension $d$)
such that for all $L \geq 16$,  all  $N$, and  all $j =1, \ldots N$, we have
\begin{equation}
\| \widehat  \Mcal_{A,k}(p) \| \leq c   M_{k,c,L}(p)
\end{equation}
for  $k= 0, \ldots, N$
and
\begin{equation}
\|  \widehat \Tcal_{A,k+1}(p)  \widehat \Mcal_{A,k}(p) \| \leq c 
\widetilde M_{k,c,L}(p)
\end{equation}
for $k = 0, \ldots, N-1$.
\end{lemma} 

\begin{proofsect}{Proof}
This is similar to Lemma~\ref{lemma:estimateMk} for the case of real symmetric $A$. 
However,  the bound $ \norm{\widetilde \Rcal(p)} \leq 1$
 for $k \leq j$ has to be replaced by
\begin{equation}
\| \widehat \Rcal_{A,k}(p)\| \leq  1 + \| \widehat\Tcal_{A,k}(p)\| \leq  1 +  c (|p| L^k)^4.
\end{equation}
This yields
\begin{equation}
\prod_{k=0}^j \| \widehat \Rcal_{A,k}(p)\| \leq   \prod_{k=0}^j  (1 +  c (|p| L^k)^4) \leq c,
\end{equation}
resulting in the additional factor $c$ in the claim.
\qed
\end{proofsect}

With the help of Lemma~\ref{lemma:sum_estimates}, we can now bound  $\norm{\Cscr_{A,k}} $ in the same way as in the proof of Theorem~\ref{THM:FRD}, starting from \eqref{E:CN+1A} for $k=N+1$ and from the equality 
\begin{equation}  
\Cscr_{A,k} =   
 (\Tscr_{A,k}  \Mscr_{A,k-1}) \Cscr_A   \Mscr_{A,k-1}'  +  \Mscr_{A,k-1} \Cscr_A ( \Tscr_{A,k}  \Mscr_{A,k-1})' 
  -   \Tscr_{A,k}  \Mscr_{A,k-1} \Cscr_A ( \Tscr_{A,k}  \Mscr_{A,k-1})' 
\end{equation}
for  $\ k=1,\dots, N$. The latter 
 follows from \eqref{E:CkA} using the equation $\Rscr_{A,k} = \id -  \Tscr_{A,k}$.

We can now use the Cauchy integral formula to control the derivatives with respect to $A$.

\begin{lemma} \label{lemma:Cauchy} Let $D = \{ z \in \C : |z| < 1 \}$.
\begin{enumerate}
\item[(i)] Suppose that  $f\colon D \to \C^{m \times m}$ is holomorphic and
\begin{equation}
\sup_{z\in D} \|f(z)\| \leq M.
\end{equation}
Then the $j$-th derivative satisfies
\begin{equation}
\|f^{(j)}(0)\|   \leq  M j!    \, \, .
\end{equation}
\item[(ii)] Suppose that $f\colon D \to \C^{m \times m}$ and $g\colon D \to \C^{m \times m}$ are
holomorphic and
\begin{equation}
\sup_{z\in D} \|f(z)\| \leq M_1, \quad \sup_{z\in D} \|g(z)\| \leq M_2. 
\end{equation}
Then the function $h(t) = f(t) g^*(t)$ is real-analytic in $(-1,1)$
and
\begin{equation}
\|h^{(j)}(0)\|   \leq  M_1 M_2  j!    \, \, .
\end{equation}
\end{enumerate}
Here $g^*(t)$ denotes the adjoint matrix of $g(t)$.
\end{lemma}

\begin{proofsect}{Proof}
Assertion (i) follows directly from the Cauchy integral formula.
To show (ii), we  note that 
$g(z) = \sum_j a_j z^j$ with $a_j \in \C^{m \times m}$. Define $G(z):= \sum_j a_j^* z^j$.
Then $G(z) = {g(z^*)}^*$. Hence $\|G(z)\| = \|g(z^*)\|$. Thus $H := f G$ is holomorphic
in $D$ and satisfies $\sup_D \|H\| \leq M_1 M_2$. Hence $H^{(k)}(0) \le k! \, M_1 M_2$.
For $t \in (-1,1)$ we have $H(t) = h(t)$ and  the assertion follows.
\qed
\end{proofsect}

\begin{proofsect}{Proof of Theorem \ref{THM:FRDfamily}}
It only remains to show the claim (iii). 
Let $A_0$ and $A_1$ as before and assume in addition that  $A_0$ and $A_1$ are  real and symmetric. 
Set
\begin{equation}
A(z) := A_0 + z A_1
\end{equation}
Then the maps
\begin{eqnarray}
z &\mapsto& \widehat \Mcal_{A(z),k}(p), \quad 
z \mapsto \widehat \Tcal_{A(z),k}(p)\widehat \Mcal_{A(z),k}(p), \quad
z \mapsto \widehat \Mcal_{A(z),k}(p) \widehat \Ccal_{A(z)}(p), \nonumber \\
z &\mapsto& \widehat \Tcal_{A(z),k}(p)\widehat \Mcal_{A(z),k}(p) \widehat \Ccal_{A(z)}(p)
\end{eqnarray}
are holomorphic in $D$.  
Moreover 
\begin{eqnarray}
\| \widehat \Mcal_{A(z),k}(p) \| &\leq&  c M_{k,c,L}(p), \nonumber \\ 
\| \widehat \Tcal_{A(z),k}(p) \widehat \Mcal_{A(z),k}(p) \| &\leq&  c \widetilde M_{k,c,L}(p),  \nonumber \\
\| \widehat \Mcal_{A(z),k}(p) \widehat \Ccal_{A(z)}(p) \|& \leq & \frac{c}{|p|^2} M_{k,c,L}(p),  \nonumber \\
\| \widehat \Tcal_{A(z),k}(p) \widehat \Mcal_{A(z),k}(p) \widehat \Ccal_{A(z)}(p)\| &\leq&  \frac{c}{|p|^2} \widetilde M_{k,c,L}(p),
\end{eqnarray}
again with the functions $ M_{k,C,L} $ and $ \widetilde{M}_{k,C,L} $ defined in  \eqref{function1} and \eqref{function2}, respectively.
Hence it follows from  Lemma \ref{lemma:Cauchy} that
\begin{equation}
\Bigl\|\frac{\d^j}{\d t^{j}}_{|t=0}   \widehat \Ccal_{A_0 + t A_1,k}(p)\Bigr\|  \leq \frac{c j!}{|p|^2}  \, (2 M_{k-1,c,L}(p) \widetilde M_{k-1,c,L}(p) +
\widetilde M^2_{k-1,c,L}(p) ) 
\end{equation}
for $k =1, \dots, N$ (with the obvious modification for $k = N+1$).
Thus    Lemma  \ref{lemma:sum_estimates}   and the estimate
 (\ref{eq:bound_inverse_Fourier}) for the inverse Fourier transform yield
 \begin{equation}  \label{eq:final}
\sup_{x \in \T_N} \bigl\|\big({\nabla}^{\aalpha}D_A^j\Ccal_{A_0,k}(x)(A_1,\ldots,  A_1)\bigr\|
\le C_{\aalpha}(d) j! L^{-(k-1)(d-2+|\aalpha|)}L^{\eta(\alpha, d)}  .
\end{equation}
Finally suppose that $\|\dot{A}\| \leq 1$ and set $A_0 = \frac{c_0}{2} \dot{A}$. 
Then the desired estimate \eqref{eq:analyticbounds} follows from \eqref{eq:final}.
\qed
\end{proofsect}

\end{document}